\documentclass[sn-mathphys,Namedate]{sn-jnl}

\usepackage{graphicx}%
\usepackage{subcaption} 
\usepackage{multirow}%
\usepackage{amsmath,amssymb,amsfonts}%
\usepackage{amsthm}%
\usepackage{mathrsfs}%
\usepackage[title]{appendix}%
\usepackage{xcolor}%
\usepackage{textcomp}%
\usepackage{manyfoot}%
\usepackage{booktabs}%
\usepackage{algorithm}%
\usepackage{algorithmicx}%
\usepackage{algpseudocode}%
\usepackage{listings}%
\usepackage{float}

\theoremstyle{thmstyleone}%
\theoremstyle{thmstyletwo}%

\theoremstyle{thmstylethree}%

\begin{document}

\title[Article Title]{Surprising Resilience of Science During a Global Pandemic: A Large-Scale Descriptive Analysis}


\author*[1]{\fnm{Kian} \sur{Ahrabian}}\email{ahrabian@isi.edu}
\equalcont{These authors contributed equally to this work.}

\author*[1]{\fnm{Casandra} \sur{Rusti}}\email{mrusti@isi.edu}
\equalcont{These authors contributed equally to this work.}

\author[2]{\fnm{Ziao} \sur{Wang}}\email{wang-za19@mails.tsinghua.edu.cn}

\author[1]{\fnm{Jay} \sur{Pujara}}\email{jpujara@isi.edu}

\author[1]{\fnm{Kristina} \sur{Lerman}}\email{lerman@isi.edu}

\affil*[1]{\orgdiv{Information Sciences Institute}, \orgname{University of Southern California}, \orgaddress{\city{Marina del Rey}, \postcode{90292}, \state{CA}, \country{USA}}}

\affil[2]{\orgname{Tsinghua University}, \orgaddress{\city{Beijing }, \country{China}}}



\abstract{
The COVID-19 pandemic profoundly impacted people globally, yet its effect on scientists and research institutions has yet to be fully examined. To address this knowledge gap, we use a newly available bibliographic dataset covering tens of millions of papers and authors to investigate changes in research activity and collaboration during this period. Employing statistical methods, we analyze the pandemic's disruptions on the participation, productivity, and collaborations of researchers at the top 1,000 institutions worldwide based on historical productivity, taking into account variables such as geography, researcher seniority and gender, and field of study. Our findings reveal an unexpected trend: research activity and output significantly increased in the early stages of the pandemic, indicating a surprising resilience in the scientific community. However, by the end of 2022, there was a notable reversion to historical trends in research participation and productivity. This reversion suggests that the initial spike in research activity was a short-lived disruption rather than a permanent shift. As such, monitoring scientific outputs in 2023 and beyond becomes crucial. There may be a delayed negative effect of the pandemic on research, given the long time horizon for many research fields and the temporary closure of wet labs. Further analysis is needed to fully comprehend the factors that underpin the resilience of scientific innovation in the face of global crises. Our study provides an initial comprehensive exploration up to the end of 2022, offering valuable insights into how the scientific community has adapted and responded over the course of the pandemic.

}




\maketitle


\section{Introduction}\label{intro}

The COVID-19 pandemic disrupted people's lives and work routines globally. Scientists and researchers at academic institutions were among those deeply affected by the pandemic due to the closure of colleges and universities, the move to online learning, and the shuttering of labs~\citep{statnews2020coronavirus}.
Studies published early in the pandemic warned of the adverse effect of the disruptions on research productivity, especially on female researchers and those with childcare responsibilities ~\citep{Giurgee2021multicountry,myers2020unequal,morgan2021unequal,gao2021potentially}.
Reductions in research productivity would decrease scientific innovation at the very time it is required to solve society's most urgent problems, including how to respond to these disruptions.
Therefore, it is vital for us to understand how these disruptions affect the scientific community and how to make research and scientific innovation more resilient.

Our study aims to analyze changes in researcher outcomes during the COVID-19 pandemic.
Specifically, it characterizes three types of researcher outcomes representing metrics of scientific success: i) participation, ii) productivity, and iii) collaboration. 
\textit{Participation} captures the number of scientists taking an active role in the scientific community, conducting research, and publishing papers to report research findings.
\textit{Productivity} encompasses the number of novel and impactful scientific discoveries resulting from research.
\textit{Collaboration} measures diverse researchers working together to create scientific advances.
We posit that a resilient scientific community will continue to increase the participation of new researchers, maintain their productivity, and support collaborative teamwork that creates scientific innovation.

To measure the researcher outcomes of participation, productivity, and collaboration, we use data from a large-scale bibliographic catalog called OpenAlex~\citep{priem2022openalex}.
OpenAlex contains information about hundreds of millions of publications and rich metadata about the time of publication, fields of study, authors and their affiliations, and references. 
We extract features from OpenAlex, such as unique authors, publication counts, and co-authorship edges, to develop several quantitative measures to capture research outcomes. By creating a time series for each of these measures, we can assess the impact of the COVID-19 pandemic on research success and determine the resilience of the scientific community.

Beyond characterizing the changes during the COVID-19 pandemic, our study explores the various factors that impact these research outcomes.
We take advantage of the globally heterogeneous pandemic responses to disentangle pattern changes in research outcomes.
During the pandemic, mitigation policies varied by country (e.g., China's ``zero COVID'' policy vs. Sweden's more hands-off approach), US state (e.g., strong vs. weak COVID-19 response), community (e.g., political climate and school closures), and even institution (e.g., remote work policy, vaccine mandates).
The pandemic also had a different impact on researchers in different fields (biology vs. computer science), different career stages (junior vs. senior researchers), and different gender~\citep{muric2020covid,myers2020unequal}.
These diverse responses to the pandemic create conditions that allow us to identify patterns affecting the resilience of scientific innovation.
To this end, we perform a temporal analysis of the correlation of institutional prestige, geographic region, seniority, gender, and fields of study on research outcomes.


Our study focuses on researchers affiliated with the \textit{top 1,000 most prolific institutions}\footnote{Our analysis includes the top 1,000 institutions with the most publications through 2020.} and provides a stratified analysis based on the authors' institution.
We hypothesize that the pandemic adversely affected researcher productivity and collaboration.
To test this hypothesis, we operationalize measures of interest using bibliometric information and analyze these variables.
Specifically, we pose the following research questions:
\begin{description}
\item[RQ1] How did researcher \textit{participation} change during the pandemic at different research institutions world-wide?
\item[RQ2] How did researcher \textit{productivity} change during the pandemic? 
\item[RQ3] How did research \textit{collaborations} change during the pandemic?
\item[RQ4] Were there systematic differences in research outcomes based on the ranking and geographic region of institutions, researcher seniority and gender, and research publication's field of study?

\end{description}

Our examination of over 25 million publications from 10 million authors at these top 1,000 institutions has yielded an unexpected insight: the COVID-19 pandemic catalyzed a period of accelerated scientific activity. This finding stands in stark contrast to several smaller-scale studies conducted in the early stages of the pandemic that identified the detrimental impact of the pandemic on scientific productivity~\citep{muric2020covid,myers2020unequal,gao2021potentially}. In this paper, we present a comprehensive analysis of the pandemic's effects on scientific research practices, enriching the understanding of resilience and adaptability within the academic sector amid a global crisis. To challenge prevailing assumptions and test assertions about the mechanisms by which the pandemic disruptions could affect scientific output, our study explores the factors shaping this period of intensified research activity. Our comprehensive analysis not only evaluates the immediate effects of the pandemic but also provides a foundation for understanding the long-term implications of such global disruptions on scientific progress.

The paper is organized as follows. In Section~\ref{related}, we present a review of relevant literature. We then detail our research methodology in Section~\ref{methods}, covering our dataset, the metrics for assessing research output and collaboration, and our analytical approach for measuring the effects of the pandemic.
In Section~\ref{methods} we present our results on the effects of the pandemic on research trends, including comprehensive heterogeneity analysis that covers factors related to the institution's geographic location, researcher seniority and gender, and work's field of study perspectives. The paper concludes in Section~\ref{s:discussion}, where we synthesize our findings and consider their broader implications, along with potential future research directions to further understand the pandemic's impact on the scientific community and resilience dynamics.


\section{Related Works}\label{related}
Scientific research requires efficient and robust infrastructure. Institutions, a category that includes universities, government labs, industrial labs, and national academies, provide this infrastructure~\citep{Taylor2019,Raan2013}. Despite the long tradition of bibliometric and science of science research~\citep{Fortunato2018}, the focus has only recently shifted from individual scientists~\citep{Sinatra2016,wang2013quantifying} and teams~\citep{Wuchty2007,milojevic2014principles} to analyzing how institutions affect researcher productivity, collaboration and impact~\citep{Way2019,Deville2014,burghardt2020heterogeneous}. Still, many gaps remain in our understanding of the role of institutions in the resilience of scientific innovation and how they facilitate and support scientific collaborations through global and local disruptions. 

Prior works tell us that the COVID-19 pandemic has taken a toll on parents, especially mothers~\citep{Giurgee2021multicountry}. The increased childcare responsibilities due to pandemic-related school closures have led to a drop in the work hours of researchers, especially among researchers with young children \citep{myers2020unequal,Morgan2021,staniscuaski2021gender}. The work-life balance challenges have affected the productivity of women researchers, with fewer women becoming involved in COVID-19 research~\citep{vincent2020monitoring}. 

One study examined gender disparities in the authorship of tens of thousands of preprints posted on \textit{biorXiv}, \textit{medrXiv}, and \textit{Springer-Nature} in the first six months of the pandemic~\citep{muric2020covid}. By using state-of-the-art gender inference techniques, they identified the gender of almost half a million authors, which enabled them to quantify changes in the productivity of different genders. While the number of papers and authors grew in absolute terms during the early stages of the pandemic, the proportion of female authors is lower than expected. The data showed that COVID-19 exacerbated the gender gap in research production, and fewer women were engaged in COVID-19-related research.

Further, a study conducted in January 2021 found that while the initial effects of the pandemic on scientists' research time had improved, there was a decline in the number of new projects being started. This decline was more significant for female scientists and those with young children and was consistent across all research fields ~\citep{gao2021potentially}.

These previous studies on the impact of the COVID-19 pandemic on scientific research were conducted soon after the onset of the pandemic, usually within six to twelve months. These studies were also limited in scope and size, typically including a small number of research scientists. They raised important questions about the duration of the pandemic's effect on scientific research and the recovery time for the scientific community. In contrast, our work addresses these limitations by observing a longer time frame post-pandemic onset. Specifically, we analyze data in six-month intervals through the end of 2022, providing a more comprehensive view of the pandemic's impact over time. Additionally, our study significantly expands the scale of analysis, encompassing over 25 million works by more than 10 million authors across 1,000 institutions. This extensive scope allows for a more nuanced understanding of the pandemic's effects across various dimensions of scientific research.


\section{Methods}\label{methods}

\noindent In this section, we describe the data used in our study, detailing how research institutions were chosen for analysis, and how we operationalize constructs measuring scientific productivity and collaboration. Further, we outline how we estimate the effects of the pandemic's disruption on these key metrics.


\subsection{Data}\label{ss:data}

Our study uses the massive bibliographic dataset OpenAlex~\citep{priem2022openalex}, which provides information about entities such as publications, authors, institutions, venues, and the relationships between them. OpenAlex gathers data from various sources, including Microsoft Academic Graph, and is updated every month. We used the dump obtained on September 5th, 2023. 

In processing the OpenAlex dataset, we implemented several filters to ensure data quality and relevance. First, we applied a DOI filter to guarantee the inclusion of only those publications with a Digital Object Identifier, ensuring data traceability and authenticity. Next, a completion filter was used, where we retained only complete tuples containing author ID, institution ID, work ID, and publication year. Authorship records missing any of these elements were discarded, though other data from the same publication could be retained. Additionally, we established a ranking filter based on the volume of publications from institutions before 2020. 
From this ranking, we selected the top 1,000 institutions for our analysis. Throughout the paper, we refer to these rankings based on publication count as \textit{institution rank}. 
The final dataset consists of more than 25 million works from 10 million authors across the top-ranked 1,000 institutions, forming the comprehensive scope of our study.
The analytical process involved grouping the data by high-level filters, such as continent, seniority, gender, and academic field, followed by a regression analysis within each group to identify trends and deviations. These are described in more detail below. 



\subsection{Constructs and Measurements}\label{ss:measures}
Below, we describe how we create constructs that measure research activity and collaboration.

\subsubsection{Research Participation}
Participation captures the number of scientists taking an active role at specific research institutions, which means conducting research and reporting findings in research publications. We operationalize \textit{participation} at a given research institution in a specific year as the number of active authors at that institution in that year, i.e., researchers with at least one associated publication in that year.

\subsubsection{Research Productivity}
Productivity encompasses the number of novel and impactful scientific discoveries resulting from research. A simple proxy for the productivity of an academic institution is the \textit{number of publications} associated with the institution. However, institutions may vary widely in size, creating a bias toward larger institutions. To address this issue, we also considered an \textit{individual publication count}, which takes the average of individual authors' publication count within the respective year across the scope of each analysis. 

\subsubsection{Research Collaborations}
We also study the collaboration networks of an institution's authors. Since authors collaborate within and across institutions, assessing collaborative behavior for a single institution can be difficult. To simplify the analysis, we consider the internal collaboration networks of authors, which only include collaboration links between authors at the same institution. Prior work has shown that internal collaborations are highly correlated with all collaborations~\citep{burghardt2021emergence}.

We construct and analyze scientific collaborations across the 1,000 institutions in our study over time. We represent collaborations as an unweighted, undirected network in which nodes are authors and edges represent co-authorship of papers published during a specific aggregation interval (six-month periods). 
We create a sequence of collaboration networks for an institution over time by considering only papers published in six-month interval periods.

For each institution in each year, we use the following measures of internal collaboration: \emph{average degree}, which captures the average number of internal collaborators of researchers at an institution; \textit{average clustering coefficient}, which is the ratio of observed triangles to possible triangles in the internal collaboration network, serving as a measure of the prevalence of tightly-knit collaborative communities; \emph{average team size}, which represents the average number of institution co-authors for a publication.  

\subsection{Heterogeneity Factors}\label{ss:heterog}
The response of researchers and institutions to pandemic disruptions may vary systematically based on their features. We explore several factors that could account for these diverse responses.

\subsubsection{Institution Ranking and Geographic Location}

To investigate these aspects, our analysis focuses on institutional-level data. We examine the top 1,000 institutions with the most publications through 2020. These institutions are categorized into five groups according to their publication volume ({rank} 1--200, 200--400, etc.). Given their global distribution, this approach facilitates an in-depth heterogeneity analysis, particularly examining how geographic region correlates with institutional responses to the pandemic. We limit geographic analysis to the level of continents, although finer-grained analysis is also feasible.
The distribution of publications in the sample across continents is 32\% Europe, 30\% North America, 29\% Asia, 4\% Oceania, and 4\% South America. Africa is excluded from this analysis due to the limited sample size of less than 350 thousand publications (about 1\% of our filtered data).

\subsubsection{Researcher Seniority and Gender}
We examine the relationship between the seniority of researchers and their temporal participation and productivity over the past two decades. In each year, we divide active researchers into three categories based on their level of experience: junior, mid-career, and senior. Junior researchers are defined as those who are publishing within five years of their first publication. Mid-career authors are those who publish within six and ten years of their first publication, while senior authors are those publishing eleven or more years since their first publication. The distribution of publications in the sample across researcher seniority is 31\% junior, 22\% mid-career, and 47\% senior.

We use the Python library gender-guesser~\citep{genderguesser} to infer the most likely gender for a researcher based on their first name. The possible gender labels for a researcher are ``male", ``female," ``mostly female," ``mostly male," ``androgynous," and ``unknown." Androgynous names are those that have an equal likelihood of being assigned to a male or female, while unknown names are those that cannot be determined as male or female because they were not found in the database used. In our analysis, we focus on two groups: the male measures represent authors identified as ``male" and ``mostly male,'' and the female measures represent authors identified as ``female" and ``mostly female''. The distribution of publications in the sample across researcher gender is 24\% female, 38\% male, 10\% androgynous, and 28\% unknown.

One limitation of this approach is that it gives a binary classification as ``male'' or ``female'', whereas gender is non-binary and may not be accurately captured by a person's name. 
Despite these limitations, we believe that analysis and discussion of gender-based impacts is an important component of understanding scientific resilience and informing scientific communities and policymakers of the potential for disruption in underrepresented communities.

\subsubsection{Field of Study}\label{ss:fields}

In associating papers with specific fields of study, we leverage the conceptual hierarchy in OpenAlex. For each paper, our approach involves propagating the scores of the assigned concepts (a small subset of the 65k concepts) until we reach the designated 19 root concepts. We then sum up the scores for these root concepts and ascertain the dominant concept for each paper using an argmax function, selecting the field associated with the highest aggregated score.

\subsection{Modeling Pandemic Disruption Effects}\label{ss:disruption}

To estimate the effect of the pandemic disruption, we adopted an approach used in previous studies (e.g., see \citep{muric2020covid}). Specifically, we performed linear regression for a construct over a time period of 10 years and used the trained model to predict the value of the construct for the next time point. We measure the deviation, i.e., the difference between the actual and predicted values of the construct. We then slide the time window over by one year and repeat the procedure. We calculate this relative deviation, i.e., deviation divided by the predicted value.  

\begin{equation}
    \text{Relative Deviation} = \frac{\text{Actual Value} - \text{Predicted Value}}{\text{Predicted Value}}
\end{equation}

We compute a construct’s mean and standard deviation across all selected institutions and plot the mean for the observed groups throughout this paper.  Focusing on 2020, this approach allows us to estimate the counterfactual: how the construct would have evolved if it were not for the pandemic. Then, a deviation of the real value from its predicted value gives the effect of the pandemic.

To increase the temporal granularity of our analysis, we partition the data into six-month periods. However, there is a notable imbalance in data, with  many papers listing January 1st as publication date in OpenAlex when the exact date is not available. As a result, the first half of each year has substantially more publications than the second half of the year. Therefore, we perform two separate regressions for each half of the year and combine the results in our plots. This results in a zig-zagged pattern in some of the figures throughout the paper. 

Our linear regression is set up as a time series using an ordinary least squares (OLS) approach. We evaluate the model's fit using $R^2$ scores, providing a quantitative measure of how well our model explains the observed variation in the data. For participation and productivity metrics, $R^2$ scores were consistently high, ranging from a minimum of 60\% to often between 75\% and 95\%. However, in the field of study analysis, we observed a wider range of $R^2$ scores, varying from 25\% to 80\% across different fields. Collaboration metrics presented a lower goodness of fit, with $R^2$ scores spanning from 10\% to 60\%, though most were above 25\%.


For \textit{Individual Productivity} we show the time-normalized percentage increase, where the normalized publication count in the year 2010 is used as a baseline measure, and each successive year shows the percentage increase or decrease in the normalized publication count.

To conduct heterogeneity analysis, we stratified data based on the factor under consideration and conducted regression analysis on each subset. For instance, to measure how gender influenced the effects of the disruption on researchers, we separated authors based on gender and performed regression analysis within females only (or males only).


\section{Results}\label{results}

We analyze the constructs of research activity we defined above to measure deviation from historical trends. We mark first half of 2020 as the start of the pandemic and measure changes in constructs with respect to this date.

\subsection{Author Participation}\label{ss:participation}
\noindent 
\begin{figure}[t]
    \begin{subfigure}{0.32\textwidth}
        \centering
        \includegraphics[width=\linewidth]{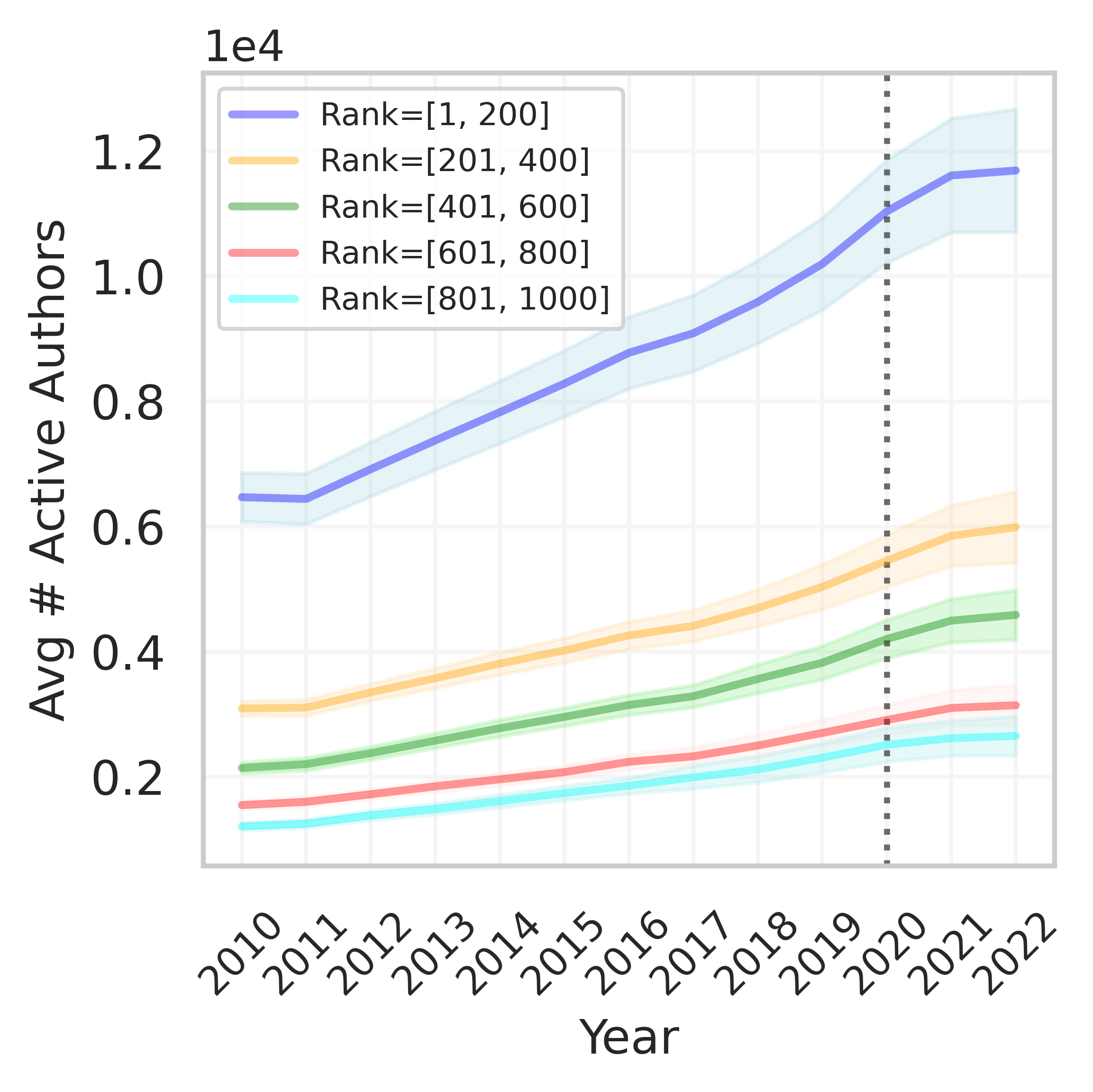}
        \caption{}
        \label{fig:rank_participation_raw}
    \end{subfigure}
    \hspace{0.03\textwidth} 
    \begin{subfigure}{0.65\textwidth}
        \centering
        \includegraphics[width=\linewidth]{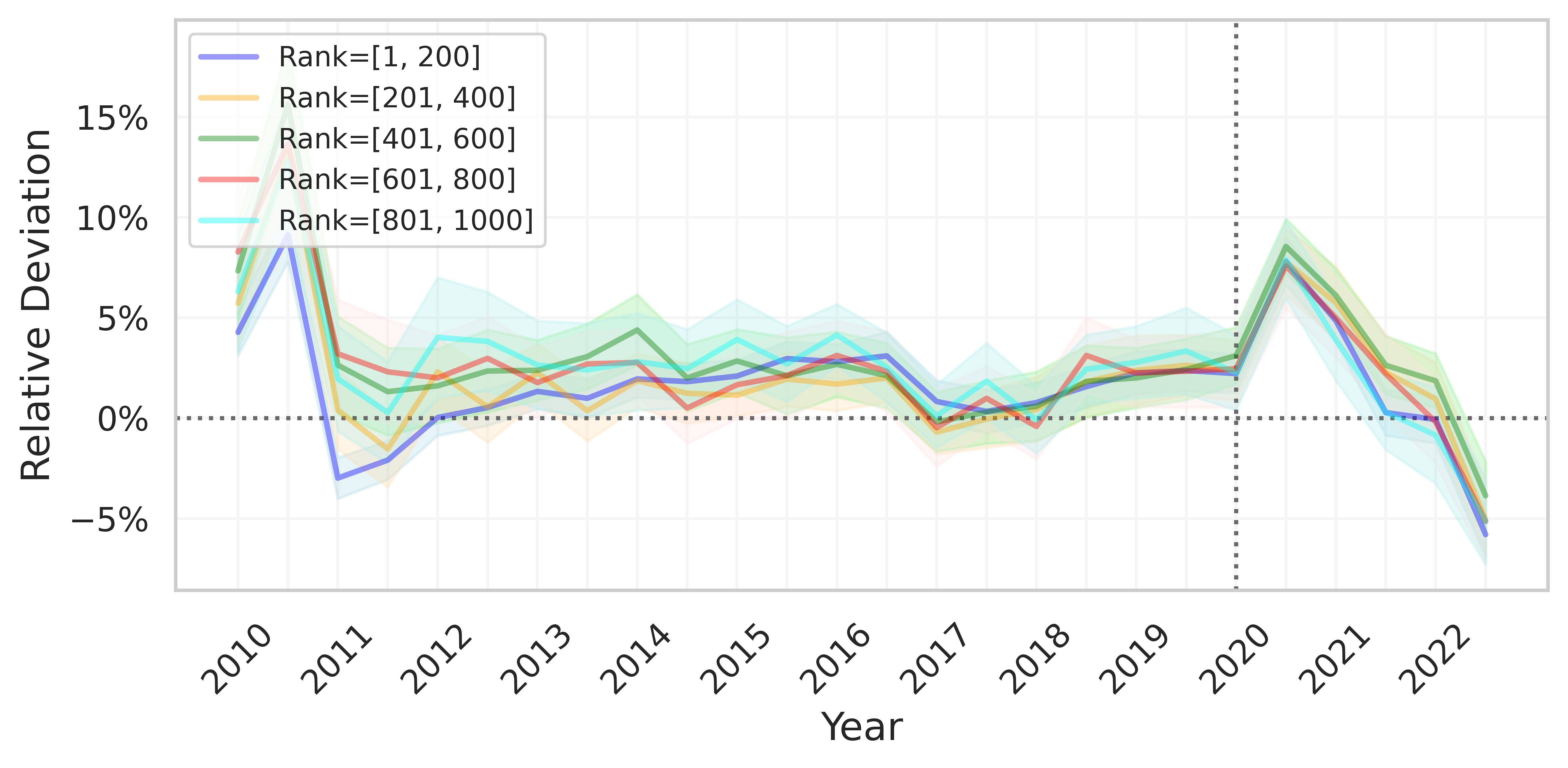}
        \caption{}
        \label{fig:rank_participation}
    \end{subfigure}
    \caption{\textit{Author Participation} in the top 1,000 research institutions grouped by \textit{Institutional Rank}, 2010-2022. (a) Raw data: Average active authors per institution, grouped by rank (in millions). (b) Relative deviation for author participation, trained on previous 10 years' data, calculated and plotted in six-month intervals.}
    \label{fig:both_rank_participation}
\end{figure}

\noindent
Figure~\ref{fig:both_rank_participation} shows that researcher participation, i.e., the average number of active authors at research institutions, varies significantly across different institutional rankings (Fig.~\ref{fig:rank_participation_raw}). However, when we plot the \textit{relative deviation} from the predicted participation (Fig.~\ref{fig:rank_participation}), 
the effect of the institutional ranking largely disappears. Surprisingly, this is true both before the pandemic and also how the institutions are impacted by the pandemic. 
Specifically, across all institutional rankings, we see an increase in \textit{author participation} early in the pandemic, reaching its peak (+10\% relative deviation) in the second half of 2020, which then reverts to pre-pandemic trends by the end of 2022. It's important to recognize that the figures show a roughly -5\% negative relative deviation in late 2022. However, this deviation is being compared to a linear model trained through the prior year, which includes the late 2020 spike. In contrast, when we examine the relative deviation versus a linear model trained on data frozen through 2019, we find that late 2022 resembles a return to pre-pandemic levels of participation. The only slight exception to this trend is among the top-ranked institutions (rank 1-200), where we observe an average decrease in author participation of -1\% in late 2022 when compared using linear regressions on the 2010-2019 data (Table~\ref{tab:rank:auth}).

The $R^2$ values of the regression are quite high, indicating a good fit of the regression models to the data. The $R^2$ values depend on institution rank: they range from approximately 85\% to 90\% for rank 1-200 to around 70\% to 75\% for rank 801-1,000. Notably, higher-ranked institutions also have narrower confidence intervals, indicating that deviations within this group fall within a tighter range. 

The consistently positive deviations suggest super-linear growth in active authors at the top 1,000 institutions over the past two decades, indicating that a non-linear predictive model might yeald more accurate predictions and higher $R^2$ values.

\subsection{Productivity}\label{ss:productivity}
\begin{figure}[t]
    \begin{subfigure}{0.32\textwidth}
        \centering
        \includegraphics[width=\linewidth]{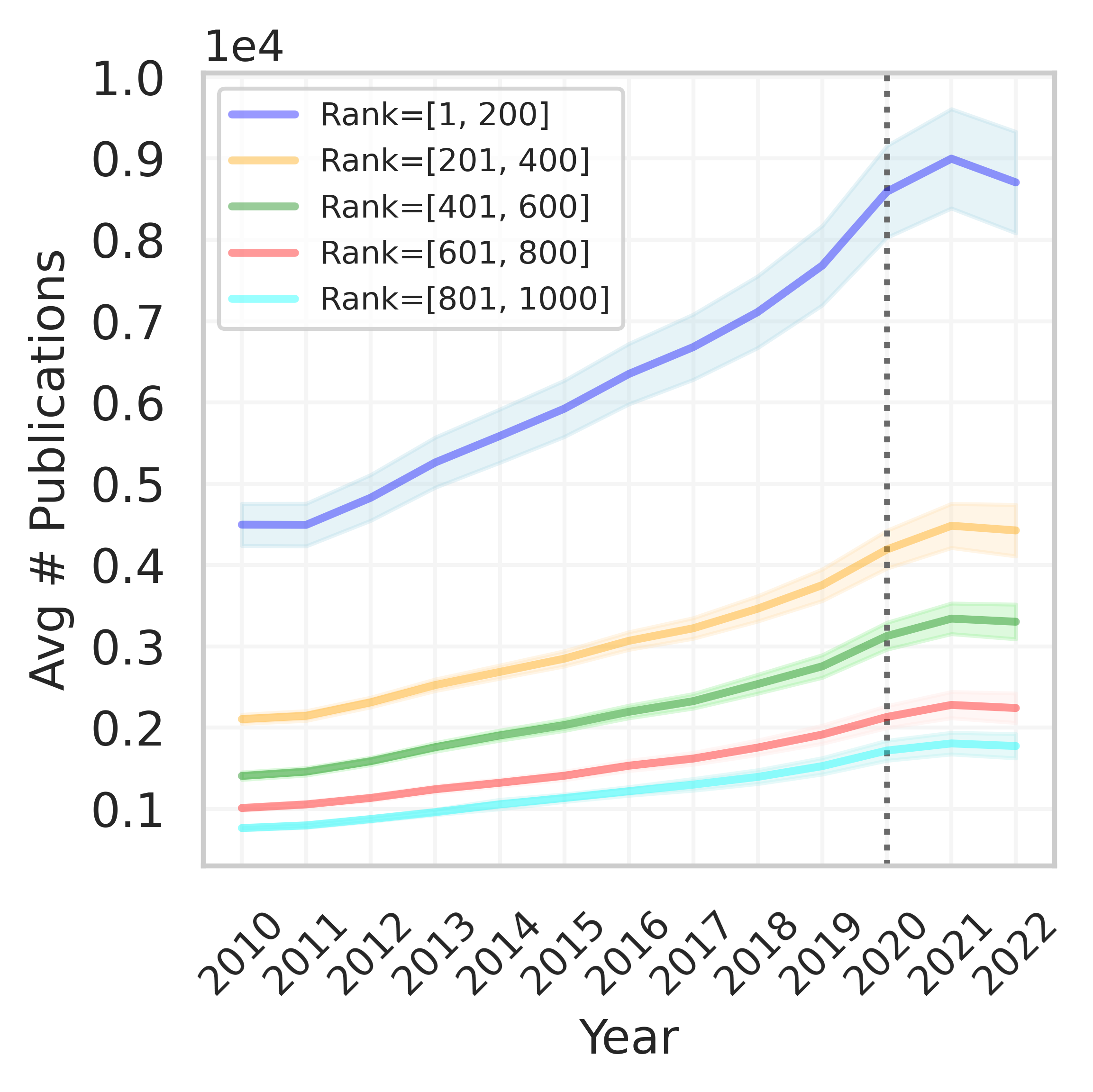}
        \caption{}
        \label{fig:rank_publications_raw}
    \end{subfigure}
    \hspace{0.03\textwidth} 
    \begin{subfigure}{0.65\textwidth}
        \centering
        \includegraphics[width=\linewidth]{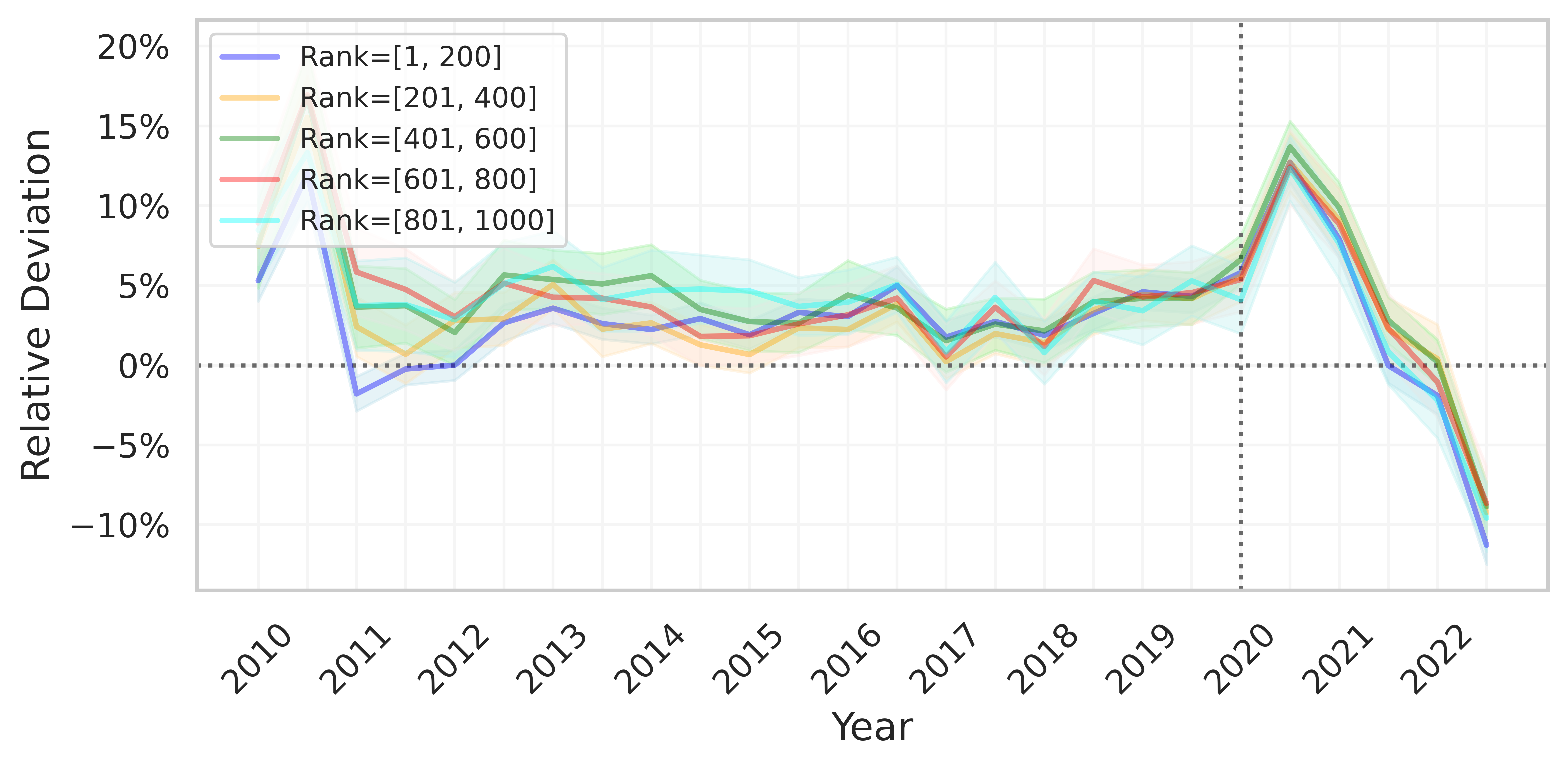}
        \caption{}
        \label{fig:rank_publications}
    \end{subfigure}
    \caption{\textit{Institutional Productivity} in the top 1,000 universities grouped by \textit{Institutional Rank}, 2010-2022. (a) Raw data: Average number of publications per institution, grouped by rank (in millions). (b) Relative deviation of average publications from values predicted by a linear model trained on previous 10 years' data, calculated and plotted in six-month intervals.}
    \label{fig:both_rank_publications}
\end{figure}

\begin{figure}[t]
    \centering
        \includegraphics[width=\textwidth]{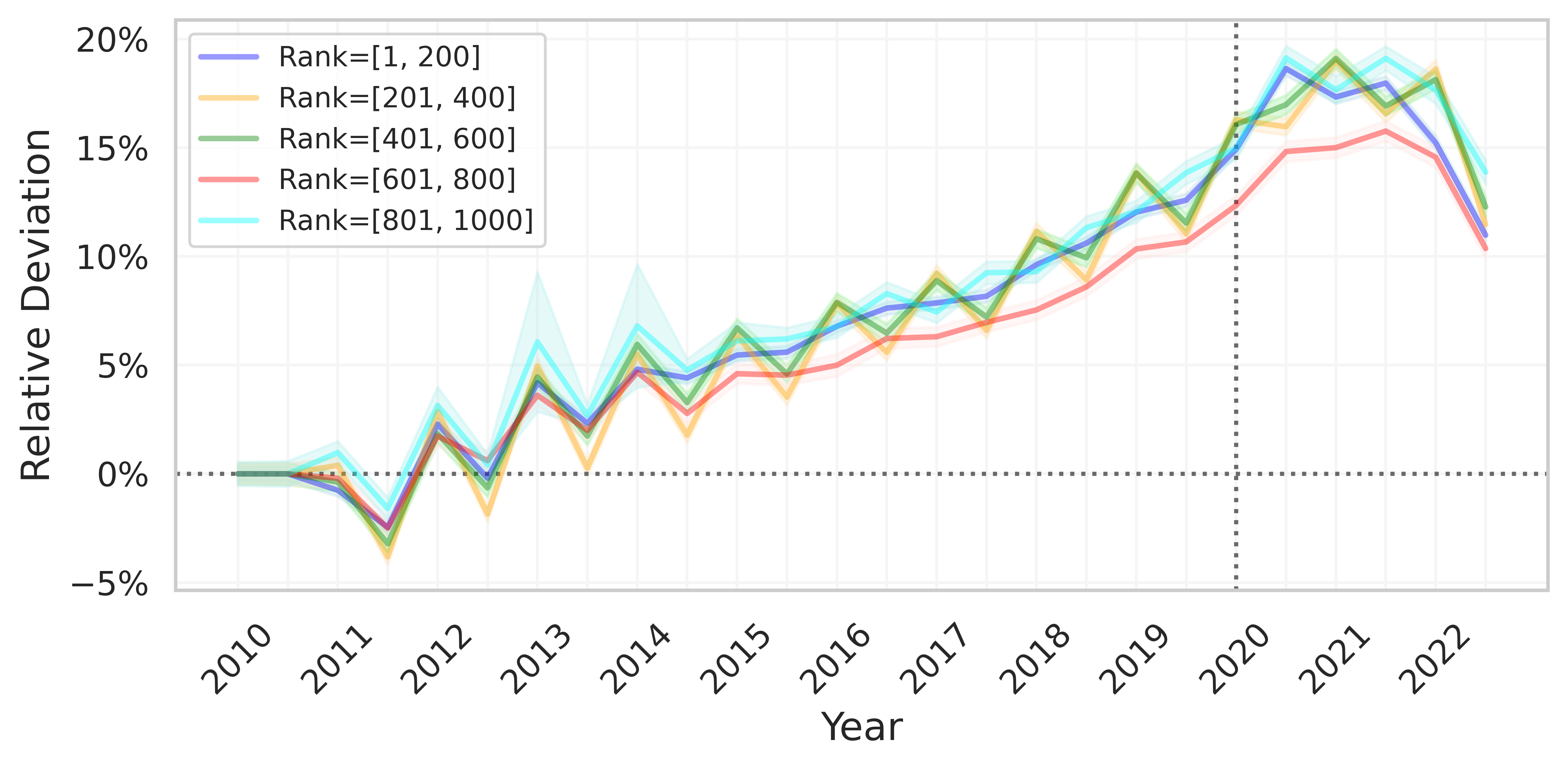}
        \caption{\textit{Individual Productivity}, relative to 2010, in the top 1,000 universities grouped by \textit{Institutional Rank}, calculated and plotted in six-month intervals between 2010-2022.}
        \label{fig:rank_productivity}
\end{figure}

\noindent 
Similar observations can be extended to \textit{institutional productivity} (Fig.~\ref{fig:both_rank_publications}). Notably, the surprising surge in author participation was associated with a significant rise in the average number of publications in late 2020 and early 2021.
Moreover, the observed 10\% increase in the number of authors (Fig.~\ref{fig:both_rank_participation}) was further amplified by an unexpected 15-20\% boost in \textit{individual productivity} (Fig.~\ref{fig:rank_productivity}). This dual increase culminated in a peak of 15\% in \textit{institutional productivity} deviation during late 2020 across all institutional rank groups in the top 1,000 institutions.

Once more, we observe a reversion to \textit{pre-pandemic} total institutional publication trends by late 2022, which corresponds to the -10\% trend in late 2022 depicted in Figure \ref{fig:rank_publications} trained on data through 2021. Notably, a similar exception is observed among the top-ranked institutions (rank 1-200), which experienced a modest decline of -4\% in late 2022 when assessed through linear regressions on the 2010-2019 dataset (Table~\ref{tab:rank:work}).

\subsection{Research Collaborations}\label{ss:colab}
\begin{figure}[t]
    \begin{subfigure}{0.32\textwidth}
        \centering
        \includegraphics[width=\linewidth]{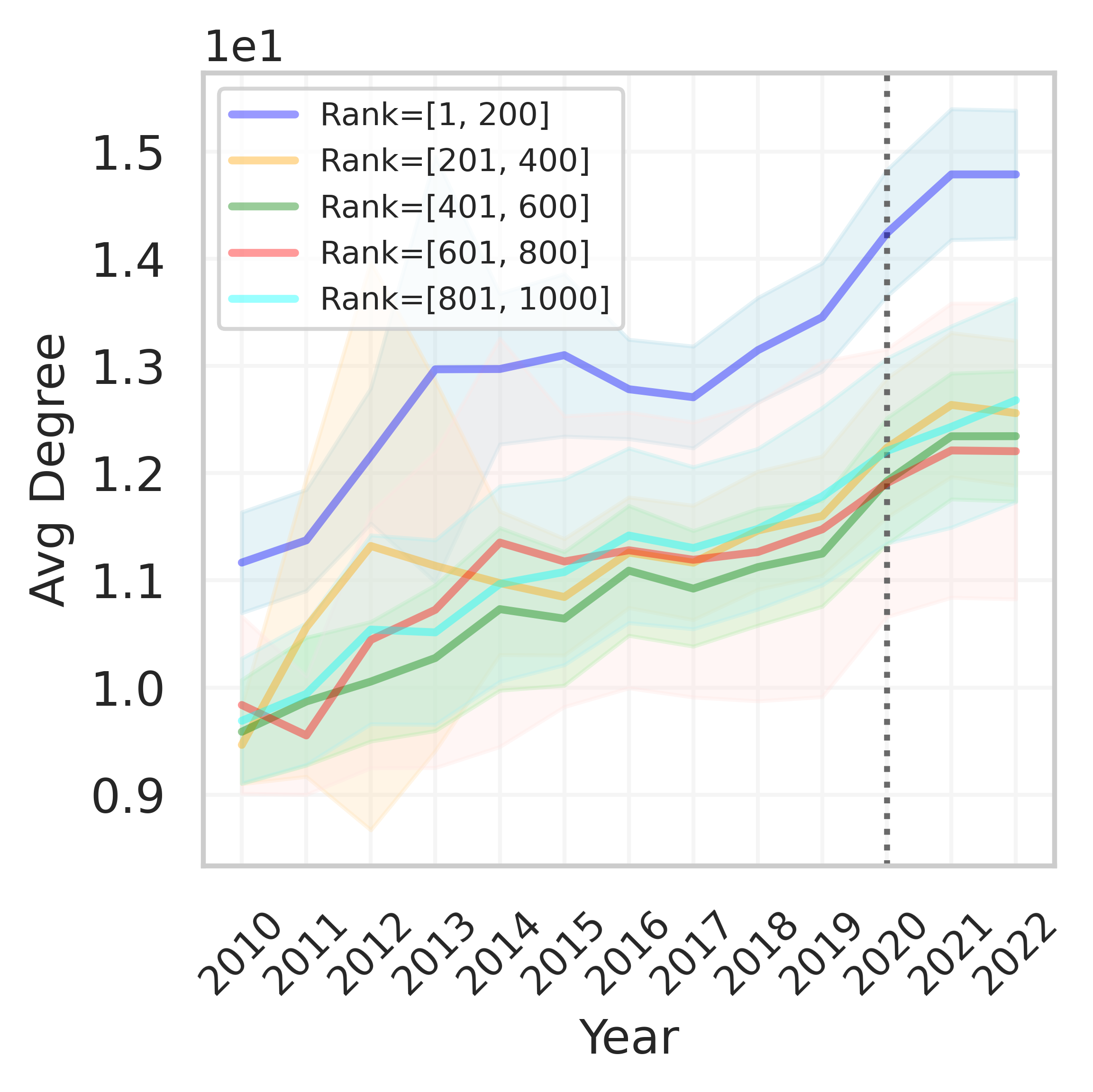}
        \caption{}
        \label{fig:rank_degree_raw}
    \end{subfigure}
    \hspace{0.03\textwidth} 
    \begin{subfigure}{0.65\textwidth}
        \centering
        \includegraphics[width=\linewidth]{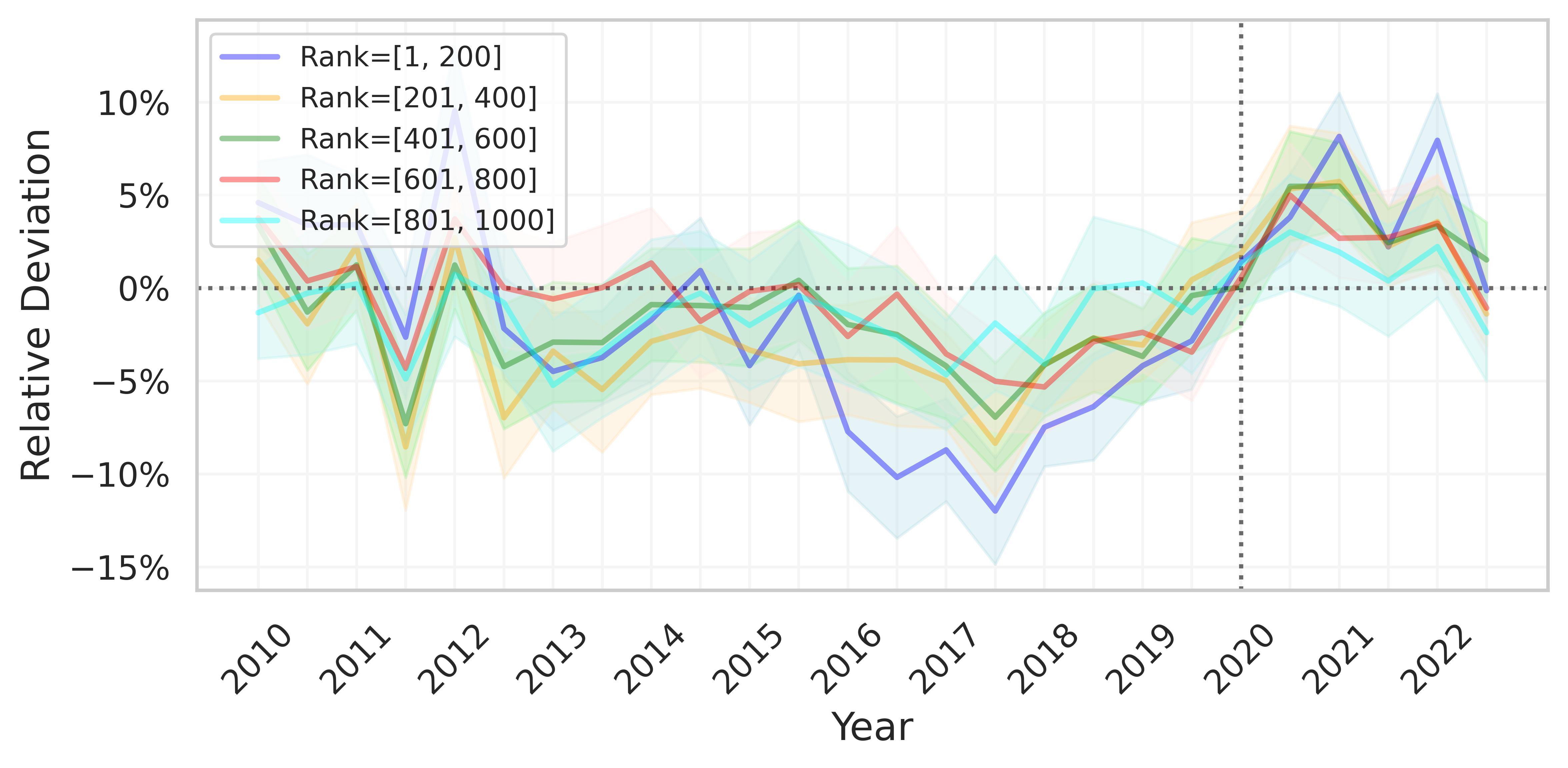}
        \caption{}
        \label{fig:rank_degree}
    \end{subfigure}
    \caption{\textit{Average number of internal collaborations} (average author degree) in the top 1,000 research institutions grouped by \textit{institutional rank}, 2010-2022. (a) Raw data: Average number of internal collaborations per institution, grouped by rank. (b) Relative deviation for average degree, trained on previous 10 years' data, calculated and plotted in six-month intervals.}
    \label{fig:both_rank_degree}
\end{figure}

The pandemic's impact on collaborations, as shown in Figures~\ref{fig:both_rank_degree}, \ref{fig:both_rank_clustercoef}, and \ref{fig:both_rank_teamsize} reveal intriguing dynamics. The \textit{average degree}, which represents the \textit{average number of internal collaborators per author}, exhibits notable shifts during the pandemic. While this metric displayed a period of negative deviation over the five years prior to the pandemic, there was a noticeable rebound during the pandemic. In late 2020 and early 2021, we observed an approximate 5\% increase in this metric. Remarkably, certain institutions within the top 200 ranks experienced a substantial surge of around 10\%, especially within the first half of 2021 and 2022. 

Surprisingly, the cohesiveness of institutional researcher networks, as illustrated by the network's \textit{clustering coefficient} and the \textit{average team size}, remained relatively stable in the face of the pandemic-induced shifts in work patterns. As evidenced in Figures~\ref{fig:both_rank_clustercoef} and \ref{fig:both_rank_teamsize}, the observed deviations, while positive, were small, amounting to less than 2\%. 

\subsection{Heterogeneity Analysis}\label{ss:heteroganalysis}
\begin{figure}[t]
        \includegraphics[width=\textwidth]{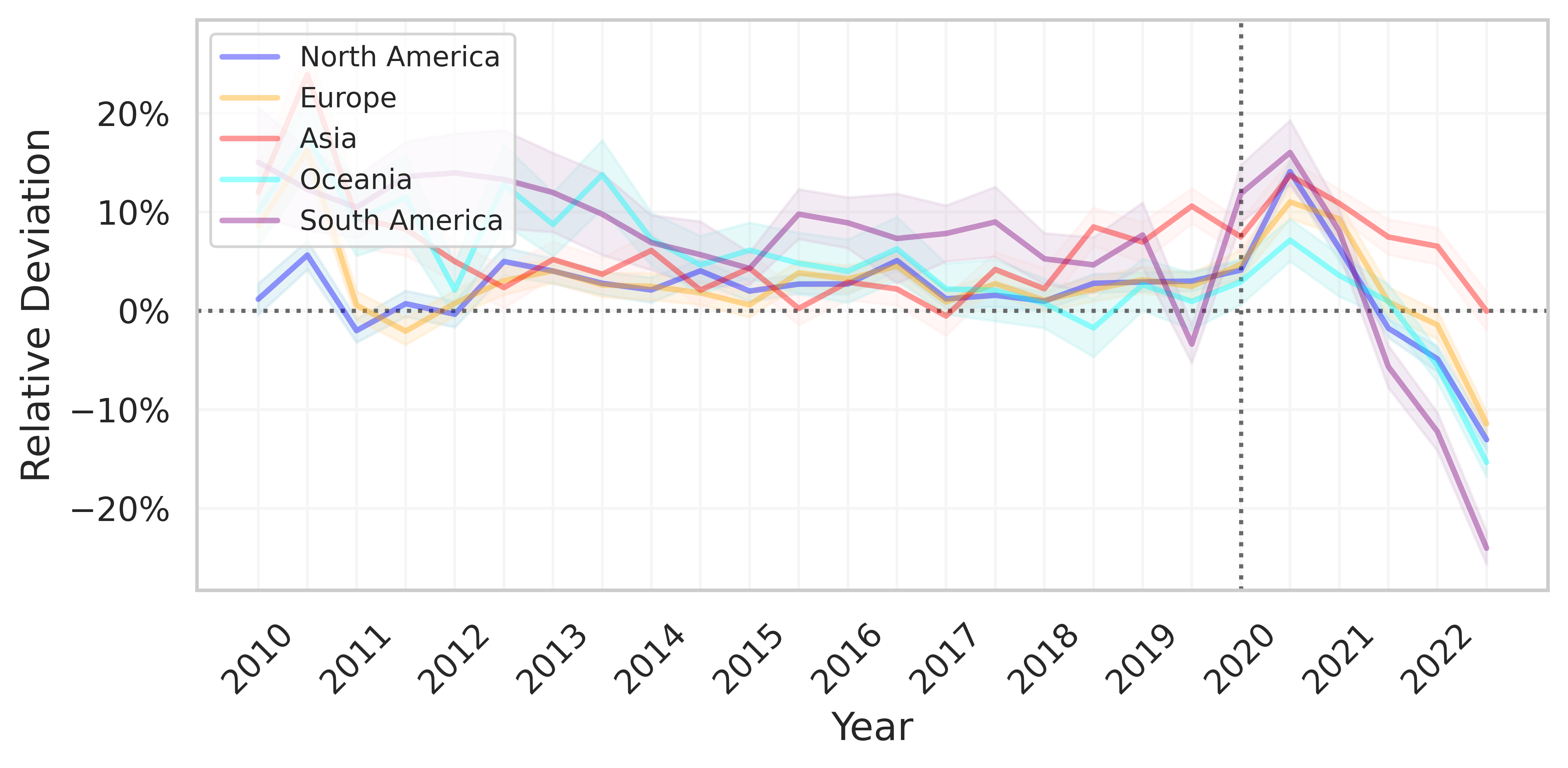}
    \caption{\textit{Institutional Productivity} in the top 1,000 universities grouped by \textit{Institutional Region}, 2010-2022. Relative deviation for average publications, trained on previous 10 years' data, calculated and plotted in six-month intervals.}
    \label{fig:continent_publications}
\end{figure}

\subsubsection{Geography}

To measure how the location of an institution impacted the effects of the pandemic disruptions on research, we group institutions by continent and carry out regression analysis separately for each continent.\footnote{We excluded Africa due to data sparseness.} 
Heterogeneity analysis of the top 1,000 institutions by continent shows distinct trends in Asia and South America during the pandemic. Asian institutions demonstrated remarkable resilience, with metrics like active authors (Fig.~\ref{fig:continent_participation}) and publications per institution (Fig.~\ref{fig:continent_publications}) aligning closely with pre-pandemic trends. This steady performance, despite the early and rigorous pandemic responses in the region, suggests a high degree of preparedness or adaptability among these institutions to the sudden shifts brought about by the pandemic.

Conversely, South American institutions appear to have faced more pronounced challenges. A -20\% relative deviation in author participation and institutional productivity by late 2022 marked a notable departure from the trend, though it translates to a modest -2\% change relative to the pre-pandemic norms (Table~\ref{tab:cont:auth}). This contrast highlights the impact of the pandemic on these institutions, potentially mirroring broader socio-economic and health-related struggles across the continent, which could have hindered their academic research and collaborations more significantly.

\subsubsection{Researcher Seniority}


\begin{figure}[t]
    \centering
        \includegraphics[width=\textwidth]{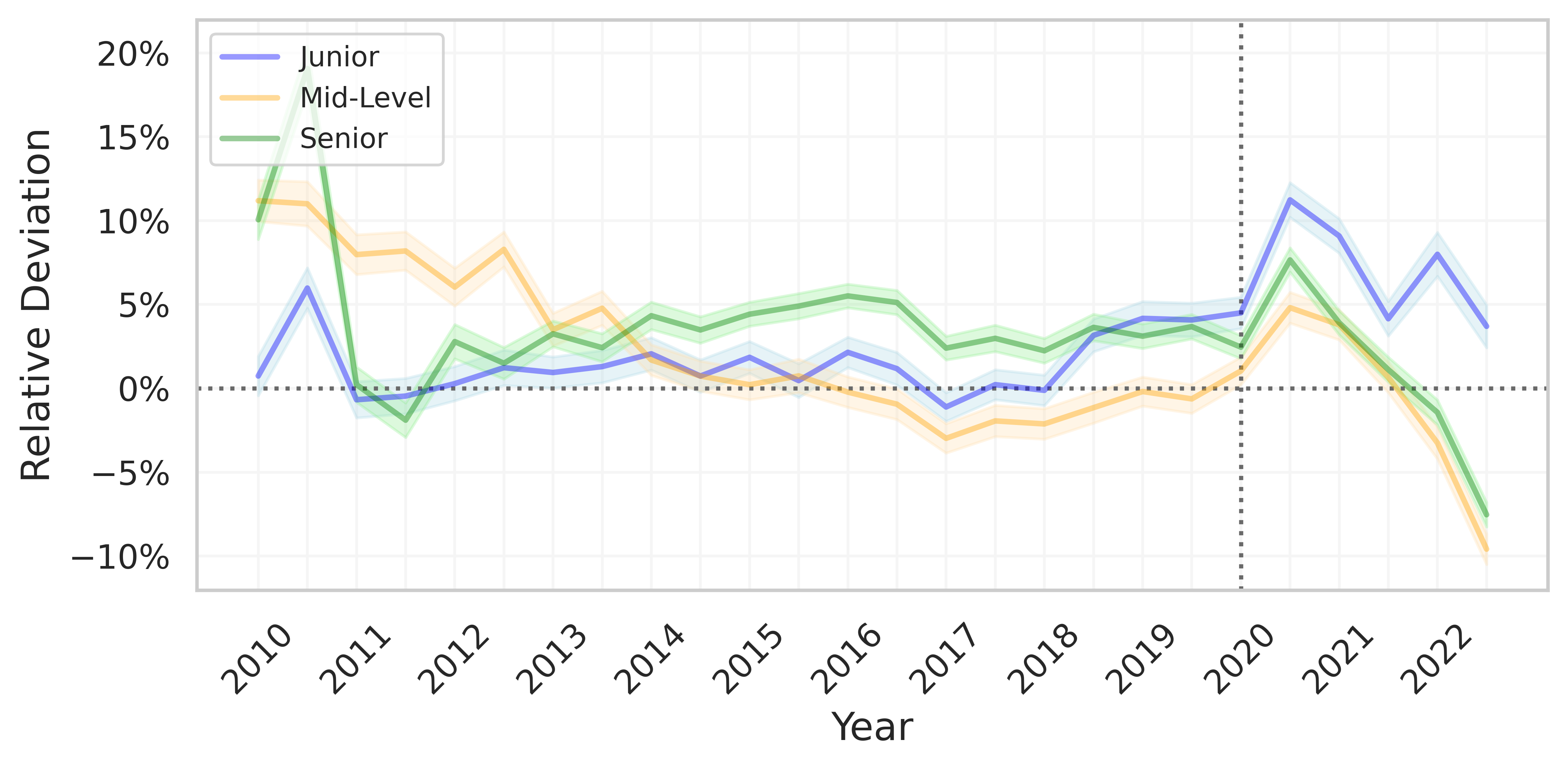}
        \caption{\textit{Researcher Seniority}: Relative deviation of {author participation} over the period from 2010--2020 at the top 1,000 universities grouped by seniority level. 
        Relative deviation for author participation, trained on previous 10 years' data, calculated and plotted in six-month intervals.}
        \label{fig:seniority_participation}
\end{figure}

\begin{figure}[t!]
    \centering
        \includegraphics[width=\textwidth]{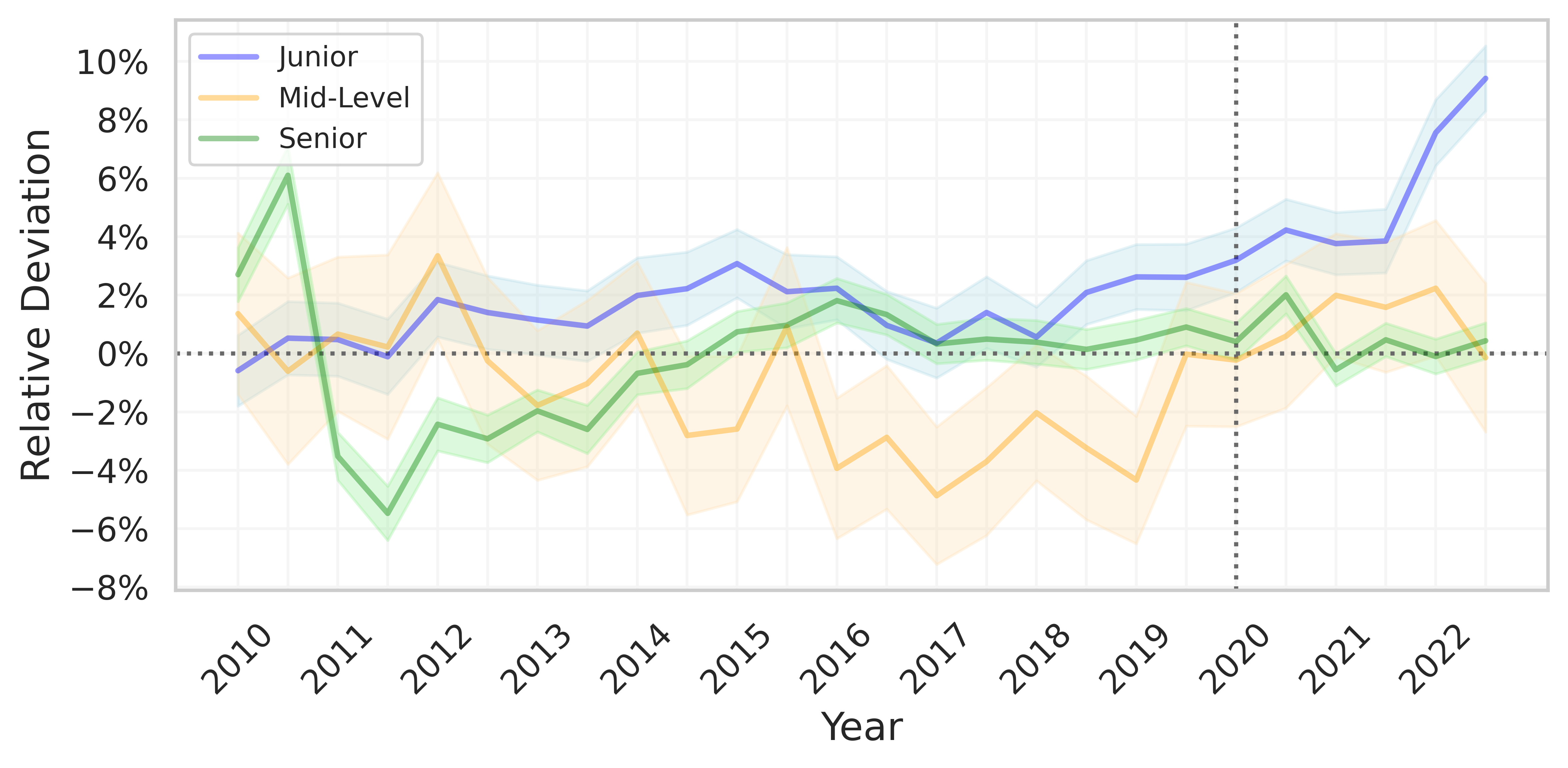}
    \caption{\textit{Average institutional internal network clustering coefficient} at the top 1,000 universities grouped by \textit{Author Seniority}, 2010-2022. Relative deviation for average institutional internal network clustering coefficient, trained on previous 10 years' data, calculated and plotted in six-month intervals.}
    \label{fig:seniority_clustercoef}
\end{figure}

In our heterogeneous analysis by seniority, we sought to examine how the pandemic's disruptions influenced scientists across different career stages (Fig.~\ref{fig:seniority_participation}). An intriguing pattern emerged during the pandemic times, characterized by a surge in participation across all seniority levels. However, this increase was most marked and sustained among junior researchers--—those with less than 5 years of publication experience. Even in late 2022, when the model accounted for an increased pace of growth in participation, junior scientists' participation rate continued to outpace the projected trend by +5\%. This figure is particularly noteworthy as it represents a +10\% deviation from pre-pandemic norms, highlighting a significant shift in engagement patterns among early-career researchers (Table~\ref{tab:sen:auth}). 
The resilience of early-career authors contrasts with smaller growth (deviation) in the participation of mid- and senior-level researchers. This could potentially be explained by greater family caregiving responsibilities of these groups of scientists.

Turning our attention to individual productivity (Fig.~\ref{fig:seniority_productivity}), we observed a reversion to pre-pandemic productivity levels by late 2022 across all seniority groups. This return to normative productivity levels underlines the transient nature of the pandemic's surprising impact on scientific output.

Lastly, our analysis uncovered a notable trend among junior-level researchers in terms of network dynamics. The clustering coefficients for this group showed a pronounced positive deviation above historical trends during the pandemic. Specifically, a +10\% relative deviation was observed in late 2022, suggesting the formation of tighter networks among junior-level researchers within institutions. This development may indicate a strategic adaptation by early-career scientists to maintain and perhaps even enhance research collaborations in response to the challenges posed by the pandemic.

\subsubsection{Researcher Gender}

\begin{figure}[t]
    \centering
        \includegraphics[width=\textwidth]{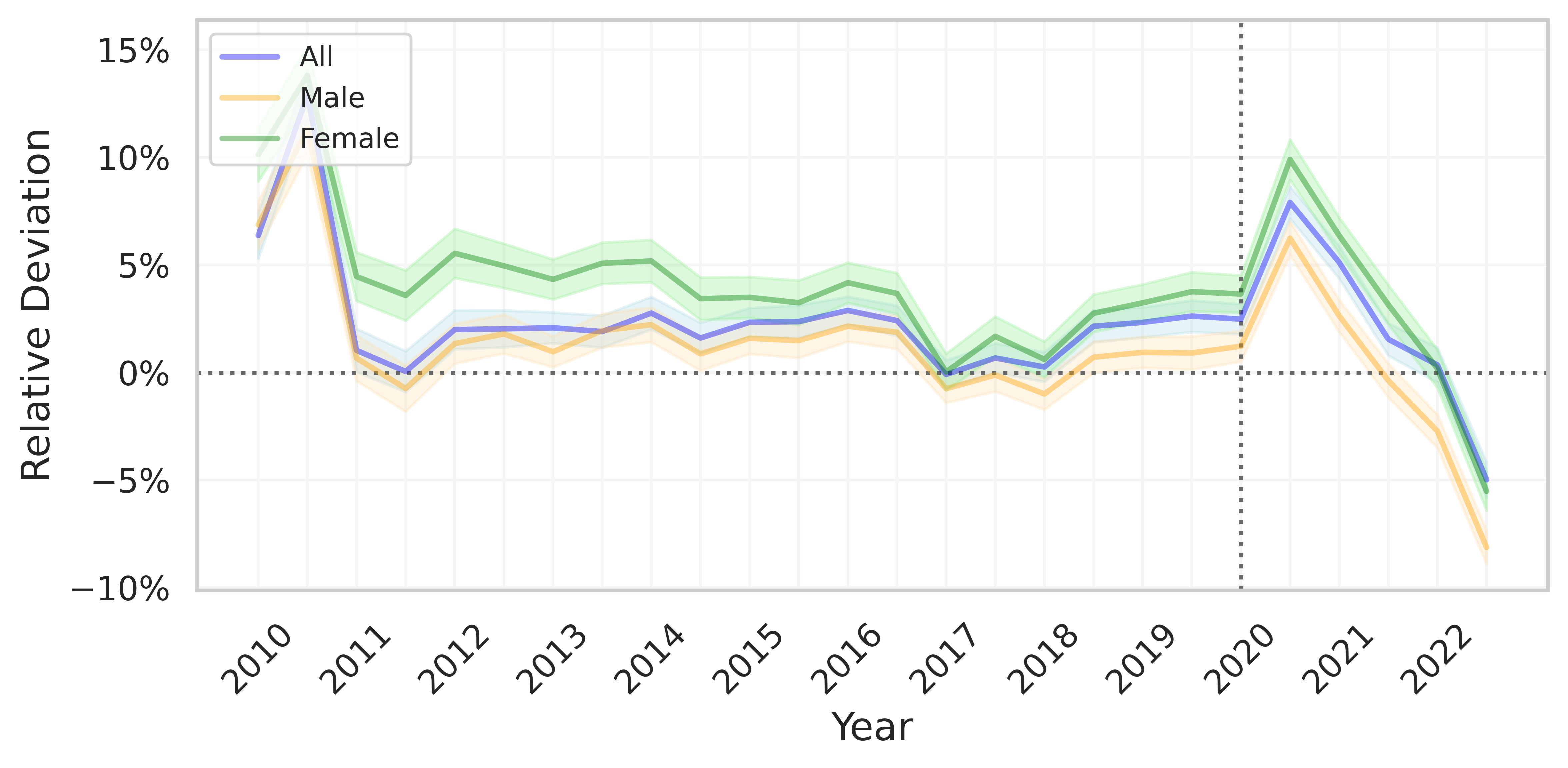}
    \caption{\textit{Author Participation} in the top 1,000 universities grouped by \textit{Author Gender}, 2010-2022. Relative deviation for author participation, trained on previous 10 years' data, calculated and plotted in six-month intervals.}
    \label{fig:gender_participation}
\end{figure}

Examining the effects of gender (Fig.~\ref{fig:gender_participation}), our findings diverge from previous studies that suggested a disproportionately negative impact of the pandemic on female scientists~\citep{gao2021potentially,muric2020covid}. Instead, we observed a remarkable surge in the participation by female researchers during the pandemic, outpacing that of male researchers, which is especially evident at its peak in late 2020. This period showed a +10\% deviation in the participation rate of female authors compared to their male counterparts up by +6\%.

These observations mark a significant development in understanding the impact of the pandemic on gender dynamics within the scientific community. They prompt further investigation into the factors contributing to this increased participation among female scientists during such a globally disruptive period.

\subsubsection{Fields of Study}

\begin{figure}[t]
        \includegraphics[width=\textwidth]{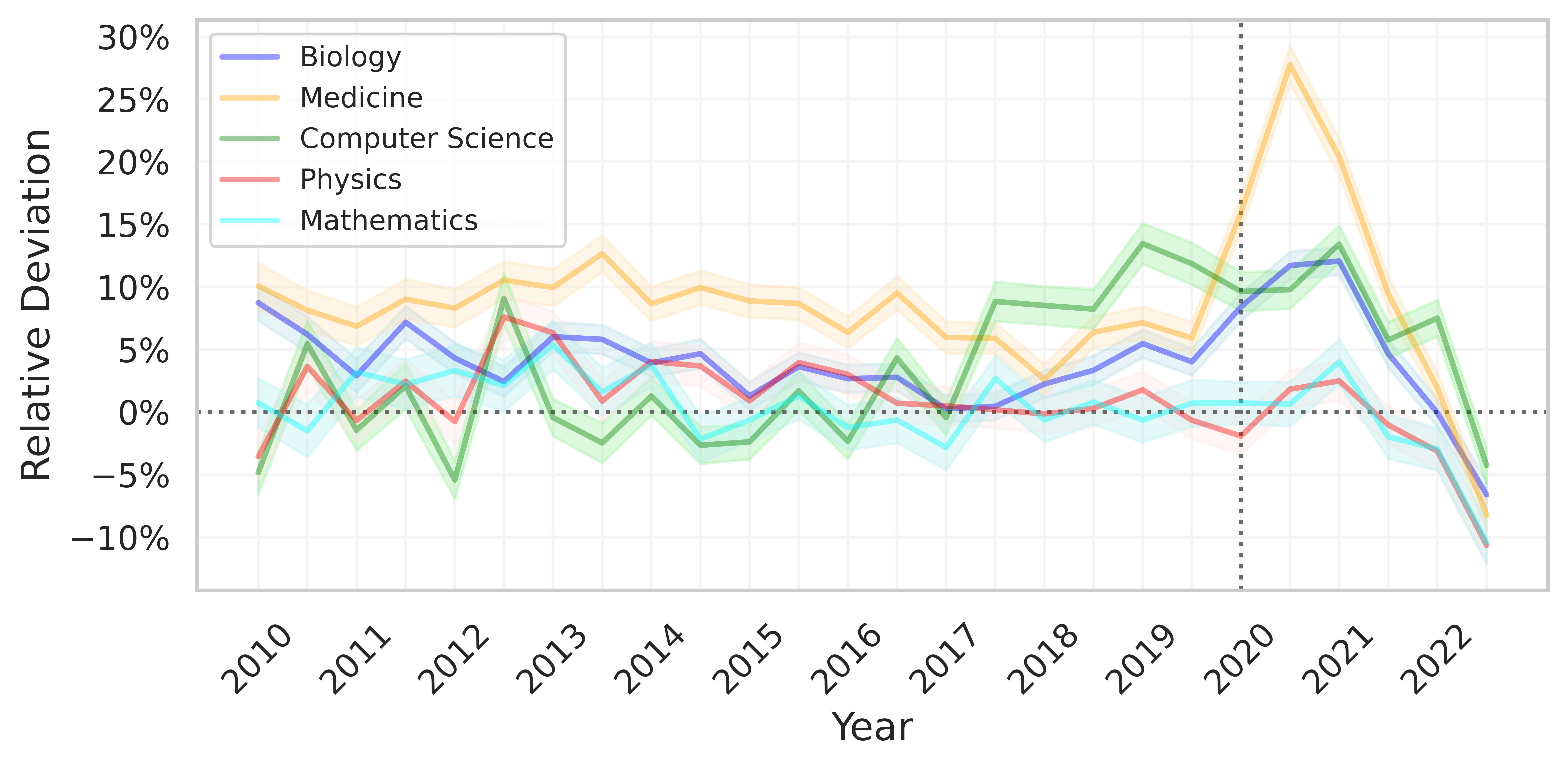}
       \caption{\textit{Institutional Productivity} in the top 1,000 universities grouped by \textit{Field of Study}, 2010-2022. Relative deviation for average publications, trained on previous 10 years' data, calculated and plotted in six-month intervals.}
        \label{fig:concept_publications}
 \end{figure}

In our study's field-specific analysis, heterogeneity in publication trends across disciplines during the pandemic is evidenced in Figures ~\ref{fig:concept_publications}, \ref{fig:concept_publications2}, and \ref{fig:concept_publications3}. Notably, the field of Medicine significantly contributed to the late 2020 publication peak, with a significant relative deviation of nearly +30\% (Fig.~\ref{fig:concept_publications}), underscoring the global focus on health research in response to the pandemic.

Conversely, other disciplines exhibited more pronounced positive deviations in the first half of 2021. However, Economics aligned with Medicine in terms of publication peak timings, showing a +10\% relative deviation in late 2020 (Fig.~\ref{fig:concept_publications2}). This trend likely reflects the substantial economic impact of the pandemic.

Psychology also displayed a notable pattern, with significant positive deviations of +10\% observed in both late 2020 and early 2021 (Fig.~\ref{fig:concept_publications3}). This dual peak suggests sustained interest in psychological aspects related to the pandemic, from immediate responses to longer-term societal impacts.

These findings highlight the differential impact of the pandemic on various academic fields, reflecting the diverse challenges and research priorities that emerged during this period.

\section{Discussion}\label{s:discussion}

Our findings offer a nuanced perspective on research trends during the pandemic, shedding light on the participation, productivity, and collaboration dynamics of scientists. The pandemic era saw a notable increase in author participation, particularly in late 2020, reflecting a significant shift in academic engagement. The reversion to pre-pandemic publication trends by late 2022, suggests that the higher research pace was not sustainable. 

We stratified research institutions based on their total publication volume before the pandemic. Despite large differences in author activity, productivity and collaboration across these institutions, their responses to the pandemic were surprisingly similar. The number of active authors and publications increased by the same relative amount at the most and least prolific institutions.

Geographically, our analysis highlighted distinct regional responses to the pandemic's disruptions. Asian institutions demonstrated remarkable resilience, with their publication and collaboration patterns closely mirroring pre-pandemic trends. This resilience may reflect effective adaptability strategies or robust institutional frameworks capable of withstanding global disruptions. In contrast, South American institutions showed more notable adverse effects (negative deviations from historical trends) in both participation and productivity. These variations suggest that regional socio-economic and public health contexts played a crucial role in shaping the academic response to the pandemic. These aspects warrant further exploration to understand the full scope of pandemic-induced changes in the academic landscape.

Another surprising finding was the more pronounced increase in research activity among early-career researchers, along with a greater participation increase among women. These results stand in contrast to papers published early in the pandemic, which noted negative impacts on women researchers. This suggests a need for intersectional analysis combining seniority and gender to potentially unravel the underlying dynamics of these gender trends.

The reduction in travel due to the pandemic may have contributed to the increased scientific output, especially among junior researchers who often have more travel engagements. Another potential contributing factor to the temporary increase in productivity could be the reallocation of time from commuting and social engagements to research activities. Rather than an overall increase in research productivity, this hypothesis could instead suggest a shift in its nature, with scientists possibly dedicating more time to manuscript preparation as opposed to experimental work.

As we examined the surprising spike in research participation and productivity, a new question emerged: does the observed negative trend in 2022 represent a return to normalcy or a depression of research activity? Our analysis, comparing 2020 through 2022 metrics against a pre-2020 linear model, indicates a reversion to pre-pandemic trends rather than a negative deviation. However, it remains to be seen whether 2023 trends will indicate a drop below historical norms or a continuation of the reversion observed in 2022.

Future research avenues are evident from our study. The application of non-linear models could provide a more accurate representation of the complex historical trends. Additionally, exploring research dynamics beyond internal institutional collaborations could unveil broader patterns of academic interactions. Further, a more elaborate field of study analysis could shed light on the specific impacts on different disciplines.





\backmatter



\bmhead{Acknowledgments}

This work was made possible by support from a Keston Exploratory Research Award.
We also thank Amazon for their generous support.
The authors acknowledge the Center for Advanced Research Computing (CARC) at the University of Southern California for providing computing resources that have contributed to the research results reported within this publication\footnote{\url{ https://carc.usc.edu}}.
Ziao Wang also acknowledges support from the Viterbi-Tsinghua University Summer Internship program.

\section*{Declarations}


\begin{itemize}
\item Funding: see Acknowledgments
\item Conflict of interest: 
The authors have no relevant financial or non-financial interests to disclose.
\item Ethics approval: Not applicable
\item Consent to participate: Not applicable
\end{itemize}


\clearpage
\begin{appendices}

\setcounter{figure}{9}
\renewcommand{\thefigure}{\arabic{figure}}

\setcounter{table}{0}
\renewcommand{\thetable}{\arabic{table}}



\section{Institutional Ranks}\label{secA1}

\begin{table}[bpth!]
\centering
\resizebox{\textwidth}{!}{
\begin{tabular}{lcccccc}
\toprule
\textbf{Rank} & \textbf{2020 H1} & \textbf{2020 H2} & \textbf{2021 H1} & \textbf{2021 H2} & \textbf{2022 H1} & \textbf{2022 H2} \\
\midrule
\textbf{1-200   } &    $0.02\pm0.07$ &    $0.08\pm0.09$ &    $0.07\pm0.10$ &    $0.04\pm0.10$ &    $0.04\pm0.12$ &   $-0.01\pm0.12$ \\
\textbf{201-400 } &    $0.02\pm0.11$ &    $0.08\pm0.12$ &    $0.07\pm0.14$ &    $0.06\pm0.15$ &    $0.05\pm0.19$ &    $0.01\pm0.18$ \\
\textbf{401-600 } &    $0.03\pm0.10$ &    $0.09\pm0.10$ &    $0.08\pm0.12$ &    $0.07\pm0.13$ &    $0.07\pm0.16$ &    $0.02\pm0.16$ \\
\textbf{601-800 } &    $0.02\pm0.12$ &    $0.07\pm0.15$ &    $0.06\pm0.17$ &    $0.06\pm0.17$ &    $0.04\pm0.21$ &    $0.00\pm0.21$ \\
\textbf{801-1000} &    $0.02\pm0.13$ &    $0.08\pm0.13$ &    $0.05\pm0.17$ &    $0.04\pm0.16$ &    $0.02\pm0.22$ &   $-0.01\pm0.20$ \\
\bottomrule
\end{tabular}
}
\caption{This figure illustrates \textit{Author Participation} across the top 1,000 research institutions, categorized by \textit{Institutional Rank}. It displays the average (and confidence interval) relative deviations in 2020 through 2022 based on an alternative linear model versus the one presented in the body of the paper, which was trained using data up to the end of 2019 (pre-pandemic). The key observation is that by late 2022, trends appear to revert to levels seen before the pandemic.}
\label{tab:rank:auth}
\end{table}


\begin{table}[bpth!]
\centering
\resizebox{\textwidth}{!}{
\begin{tabular}{lcccccc}
\toprule
\textbf{Rank} & \textbf{2020 H1} & \textbf{2020 H2} & \textbf{2021 H1} & \textbf{2021 H2} & \textbf{2022 H1} & \textbf{2022 H2} \\
\midrule
\textbf{1-200   } &    $0.06\pm0.07$ &    $0.13\pm0.09$ &    $0.11\pm0.11$ &    $0.06\pm0.10$ &    $0.05\pm0.13$ &   $-0.04\pm0.13$ \\
\textbf{201-400 } &    $0.05\pm0.11$ &    $0.13\pm0.13$ &    $0.12\pm0.16$ &    $0.09\pm0.17$ &    $0.07\pm0.20$ &   $-0.01\pm0.20$ \\
\textbf{401-600 } &    $0.07\pm0.11$ &    $0.14\pm0.12$ &    $0.13\pm0.14$ &    $0.10\pm0.13$ &    $0.08\pm0.17$ &   $-0.00\pm0.17$ \\
\textbf{601-800 } &    $0.05\pm0.14$ &    $0.12\pm0.17$ &    $0.12\pm0.19$ &    $0.09\pm0.20$ &    $0.06\pm0.24$ &   $-0.01\pm0.25$ \\
\textbf{801-1000} &    $0.04\pm0.14$ &    $0.12\pm0.15$ &    $0.10\pm0.19$ &    $0.07\pm0.18$ &    $0.04\pm0.24$ &   $-0.03\pm0.22$ \\
\bottomrule
\end{tabular}
}
\caption{This figure illustrates \textit{Institutional Productivity} across the top 1,000 research institutions, categorized by \textit{Institutional Rank}. It displays the average (and confidence interval) relative deviations in 2020 through 2022 based on a linear model trained using data up to the end of 2019.}
\label{tab:rank:work}
\end{table}

\begin{figure}[bpth!]
    \begin{subfigure}{0.32\textwidth}
        \centering
        \includegraphics[width=\linewidth]{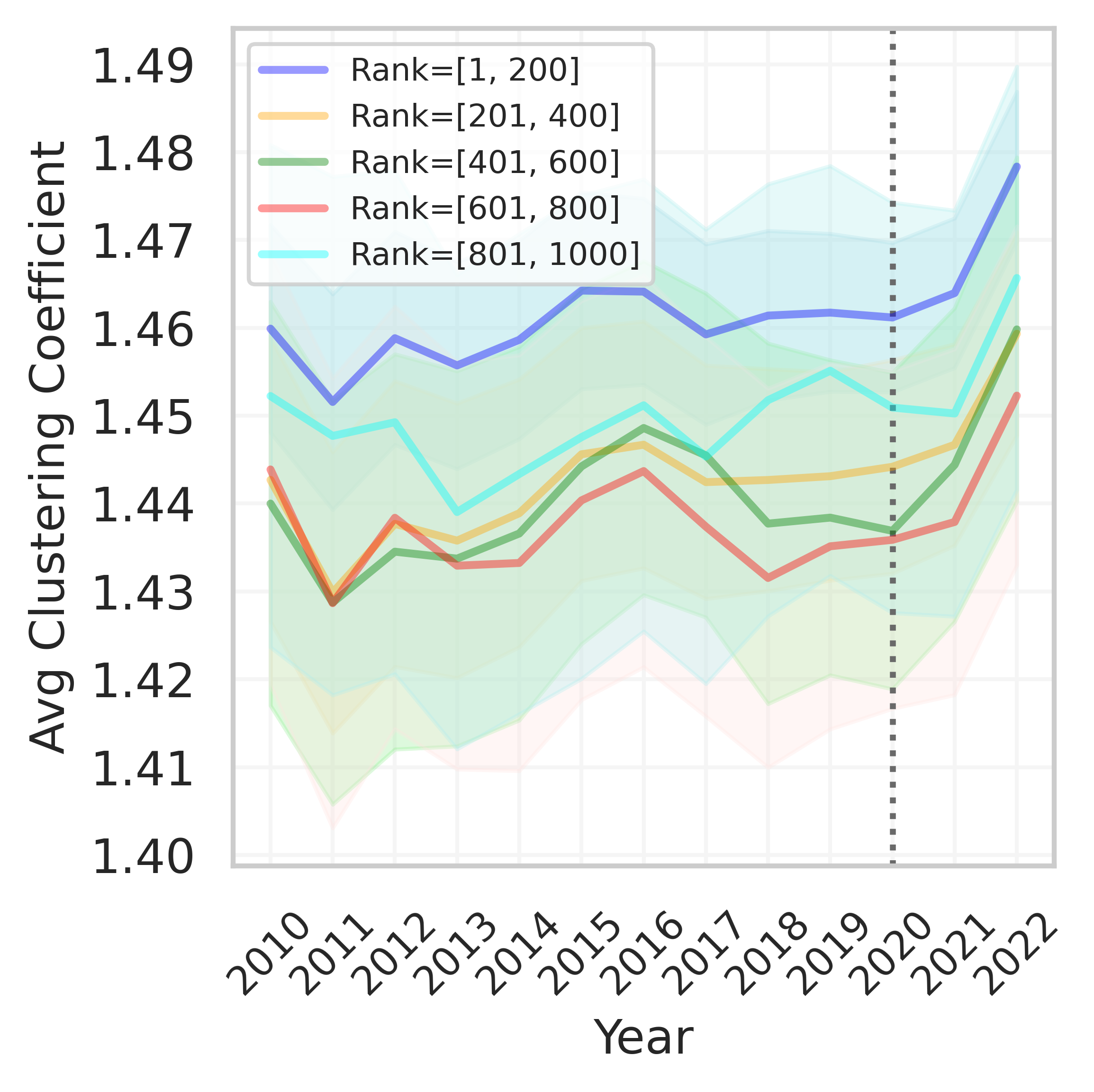}
        \caption{}
        \label{fig:rank_clustercoef_raw}
    \end{subfigure}
    \hspace{0.03\textwidth} 
    \begin{subfigure}{0.65\textwidth}
        \centering
        \includegraphics[width=\linewidth]{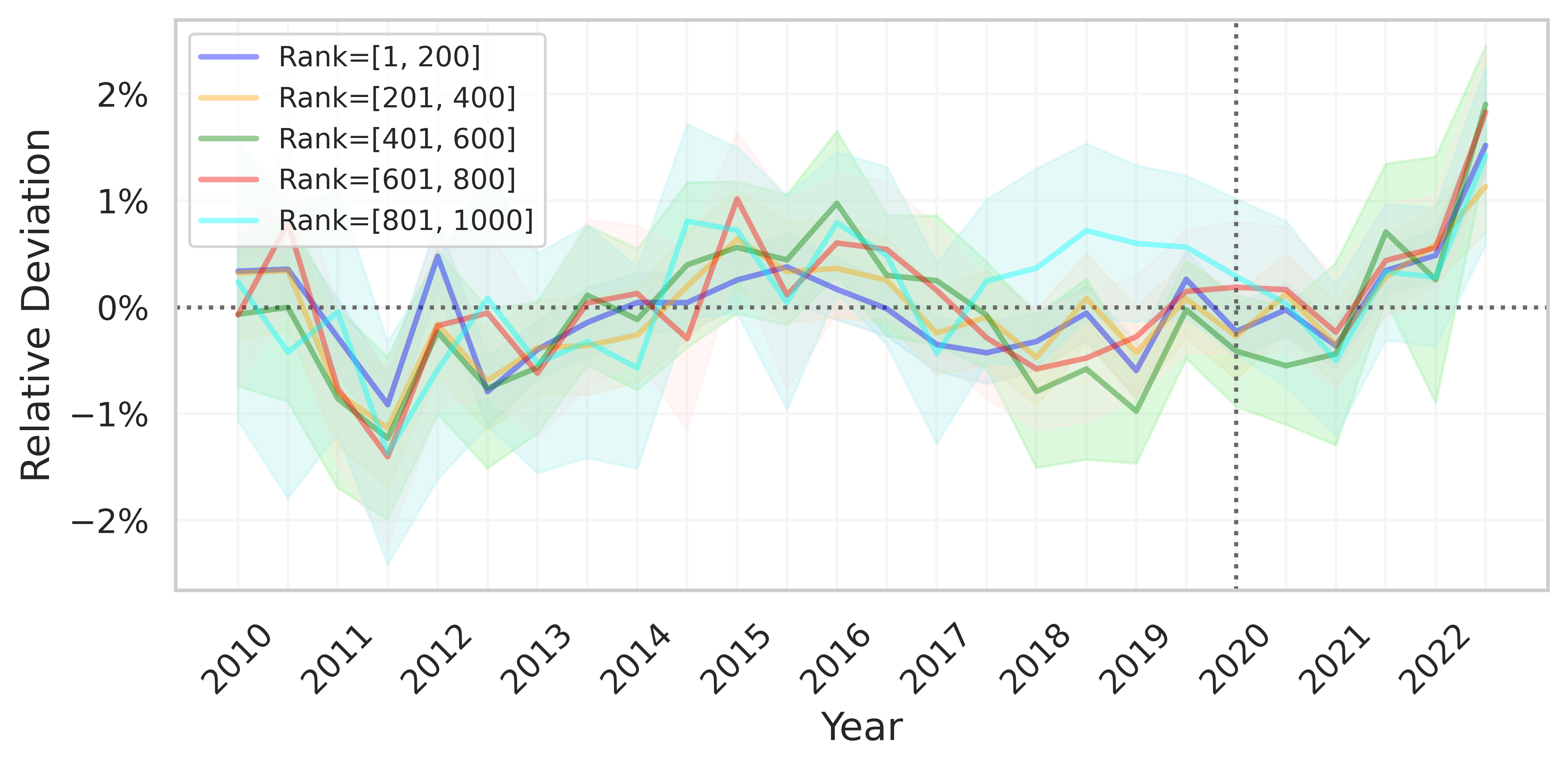}
        \caption{}
        \label{fig:rank_clustercoef}
    \end{subfigure}
    \caption{\textit{Average institutional internal network clustering coefficient} at the top 1,000 universities grouped by \textit{Institutional Rank}, 2010-2022. (a) Raw data: Average institutional internal network clustering coefficient, grouped by rank. (b) Relative deviation for average institutional internal network clustering coefficient, trained on previous 10 years' data, calculated and plotted in six-month intervals.}
    \label{fig:both_rank_clustercoef}
\end{figure}

\begin{figure}[bpth!]
    \begin{subfigure}{0.32\textwidth}
        \centering
        \includegraphics[width=\linewidth]{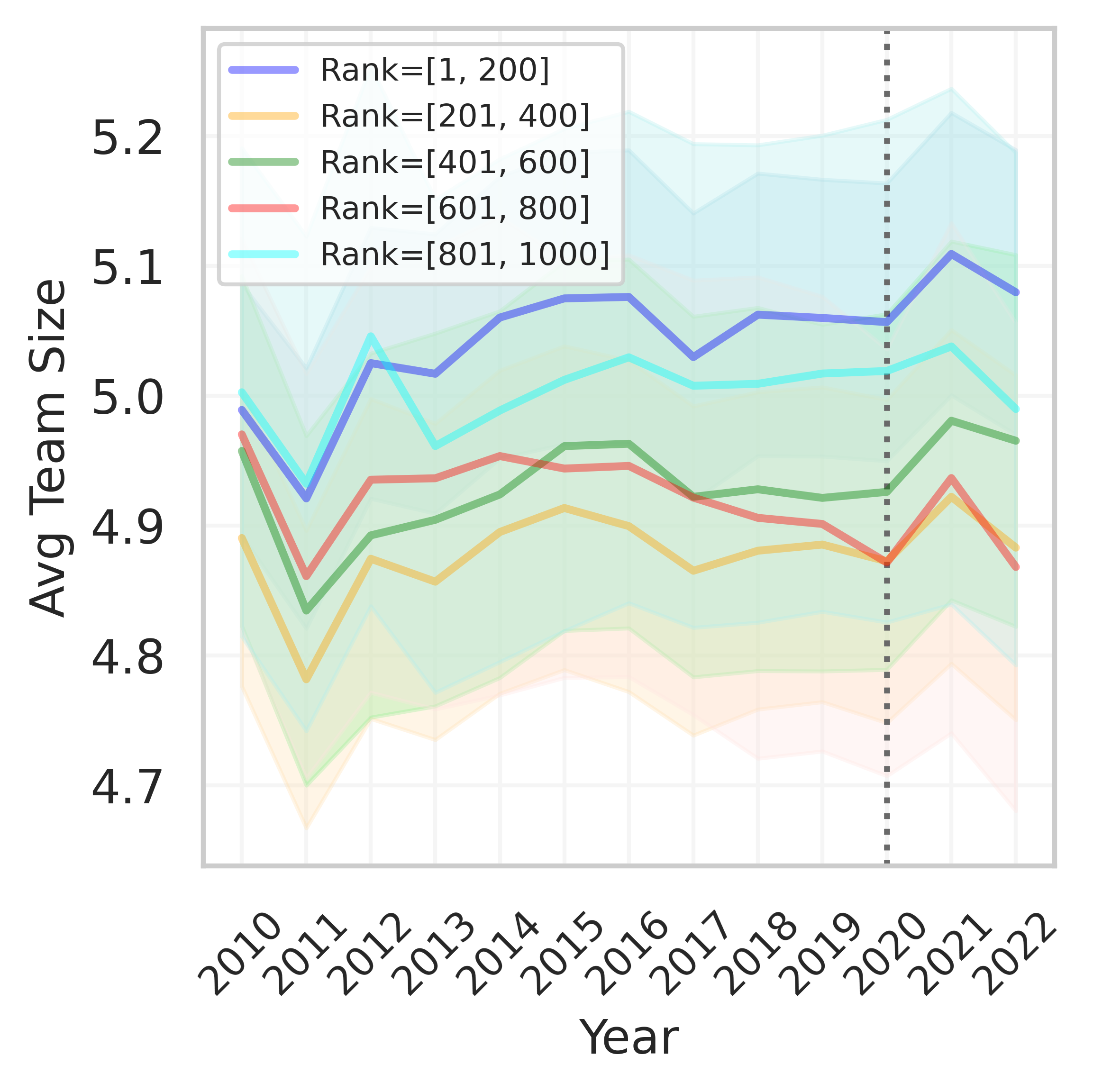}
        \caption{}
        \label{fig:rank_teamsize_raw}
    \end{subfigure}
    \hspace{0.03\textwidth} 
    \begin{subfigure}{0.65\textwidth}
        \centering
        \includegraphics[width=\linewidth]{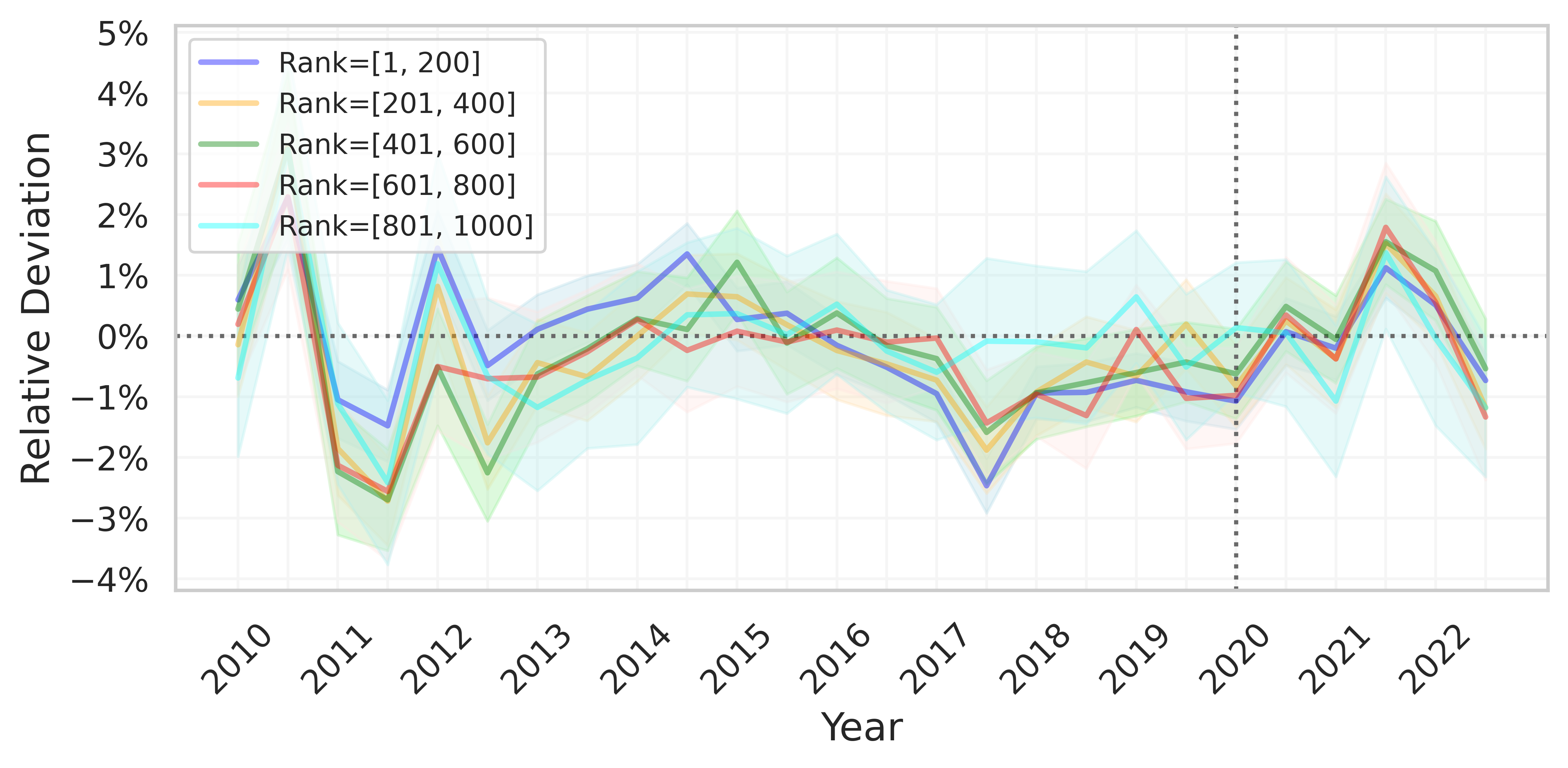}
        \caption{}
        \label{fig:rank_teamsize}
    \end{subfigure}
    \caption{\textit{Average internal team size} at the top 1,000 universities grouped by \textit{Institutional Rank}, 2010-2022. (a) Raw data: Average internal team size per institution, grouped by rank. (b) Relative deviation for average team size, trained on previous 10 years' data, calculated and plotted in six-month intervals.}
    \label{fig:both_rank_teamsize}
\end{figure}


\begin{table}[bpth!]
\centering
\resizebox{\textwidth}{!}{
\begin{tabular}{lcccccc}
\toprule
\textbf{Rank} & \textbf{2020 H1} & \textbf{2020 H2} & \textbf{2021 H1} & \textbf{2021 H2} & \textbf{2022 H1} & \textbf{2022 H2} \\
\midrule
\textbf{1-200   } &    $0.01\pm0.12$ &    $0.03\pm0.16$ &    $0.05\pm0.14$ &    $0.05\pm0.18$ &    $0.06\pm0.16$ &    $0.03\pm0.20$ \\
\textbf{201-400 } &    $0.01\pm0.15$ &    $0.04\pm0.20$ &    $0.04\pm0.18$ &    $0.04\pm0.20$ &    $0.03\pm0.17$ &    $0.02\pm0.20$ \\
\textbf{401-600 } &    $0.00\pm0.14$ &    $0.05\pm0.18$ &    $0.04\pm0.15$ &    $0.04\pm0.16$ &    $0.04\pm0.15$ &    $0.03\pm0.18$ \\
\textbf{601-800 } &    $0.00\pm0.12$ &    $0.04\pm0.17$ &    $0.02\pm0.15$ &    $0.04\pm0.18$ &    $0.03\pm0.18$ &    $0.02\pm0.20$ \\
\textbf{801-1000} &    $0.00\pm0.14$ &    $0.03\pm0.19$ &    $0.02\pm0.19$ &    $0.02\pm0.21$ &    $0.02\pm0.19$ &    $0.00\pm0.21$ \\
\bottomrule
\end{tabular}
}
\caption{This figure illustrates \textit{Average number of internal collaborations} across the top 1,000 research institutions, categorized by \textit{Institutional Rank}. It displays the average (and confidence interval) relative deviations in 2020 through 2022 based on a linear model trained using data up to the end of 2019.}
\label{tab:rank:degs}
\end{table}


\begin{table}[bpth!]
\centering
\resizebox{\textwidth}{!}{
\begin{tabular}{lcccccc}
\toprule
\textbf{Rank} & \textbf{2020 H1} & \textbf{2020 H2} & \textbf{2021 H1} & \textbf{2021 H2} & \textbf{2022 H1} & \textbf{2022 H2} \\
\midrule
\textbf{1-200   } &   $-0.00\pm0.02$ &   $-0.00\pm0.02$ &   $-0.00\pm0.02$ &    $0.00\pm0.02$ &    $0.00\pm0.02$ &    $0.02\pm0.02$ \\
\textbf{201-400 } &   $-0.00\pm0.03$ &    $0.00\pm0.03$ &   $-0.00\pm0.03$ &    $0.01\pm0.03$ &    $0.00\pm0.04$ &    $0.01\pm0.04$ \\
\textbf{401-600 } &   $-0.00\pm0.04$ &   $-0.01\pm0.04$ &   $-0.01\pm0.06$ &    $0.01\pm0.04$ &   $-0.00\pm0.09$ &    $0.02\pm0.04$ \\
\textbf{601-800 } &    $0.00\pm0.04$ &    $0.00\pm0.04$ &   $-0.00\pm0.04$ &    $0.01\pm0.04$ &    $0.01\pm0.05$ &    $0.02\pm0.05$ \\
\textbf{801-1000} &    $0.00\pm0.05$ &    $0.00\pm0.05$ &   $-0.00\pm0.06$ &    $0.00\pm0.05$ &    $0.01\pm0.06$ &    $0.02\pm0.06$ \\
\bottomrule
\end{tabular}
}
\caption{This figure illustrates \textit{Average institutional internal network clustering coefficient} across the top 1,000 research institutions, categorized by \textit{Institutional Rank}. It displays the average (and confidence interval) relative deviations in 2020 through 2022 based on a linear model trained using data up to the end of 2019.}
\label{tab:rank:clus}
\end{table}


\begin{table}[bpth!]
\centering
\resizebox{\textwidth}{!}{
\begin{tabular}{lcccccc}
\toprule
\textbf{Rank} & \textbf{2020 H1} & \textbf{2020 H2} & \textbf{2021 H1} & \textbf{2021 H2} & \textbf{2022 H1} & \textbf{2022 H2} \\
\midrule
\textbf{1-200   } &   $-0.01\pm0.03$ &    $0.00\pm0.04$ &   $-0.01\pm0.04$ &    $0.01\pm0.05$ &   $-0.00\pm0.05$ &   $-0.00\pm0.05$ \\
\textbf{201-400 } &   $-0.01\pm0.05$ &    $0.00\pm0.05$ &   $-0.01\pm0.07$ &    $0.02\pm0.07$ &   $-0.00\pm0.07$ &   $-0.00\pm0.07$ \\
\textbf{401-600 } &   $-0.01\pm0.05$ &    $0.00\pm0.05$ &   $-0.00\pm0.06$ &    $0.02\pm0.06$ &    $0.01\pm0.08$ &    $0.01\pm0.08$ \\
\textbf{601-800 } &   $-0.01\pm0.05$ &    $0.00\pm0.06$ &   $-0.01\pm0.07$ &    $0.02\pm0.08$ &   $-0.00\pm0.08$ &   $-0.00\pm0.10$ \\
\textbf{801-1000} &    $0.00\pm0.07$ &    $0.00\pm0.08$ &   $-0.01\pm0.09$ &    $0.02\pm0.09$ &   $-0.00\pm0.12$ &   $-0.00\pm0.10$ \\
\bottomrule
\end{tabular}
}
\caption{This figure illustrates \textit{Average internal team size} across the top 1,000 research institutions, categorized by \textit{Institutional Rank}. It displays the average (and confidence interval) relative deviations in 2020 through 2022 based on a linear model trained using data up to the end of 2019.}
\label{tab:rank:tmsz}
\end{table}

\newpage
\section{Institutional Regions}\label{secA2}

\begin{figure}[bpth!]
    \centering
    \begin{minipage}{0.35\textwidth}
        \caption{\textit{Author Participation} in the top 1,000 universities grouped by \textit{Institutional Region}, 2010-2022. Relative deviation for author participation, trained on previous 10 years' data, calculated and plotted in six-month intervals.}
        \label{fig:continent_participation}
    \end{minipage}%
    \hfill
    \begin{minipage}{0.63\textwidth}
        \includegraphics[width=\textwidth]{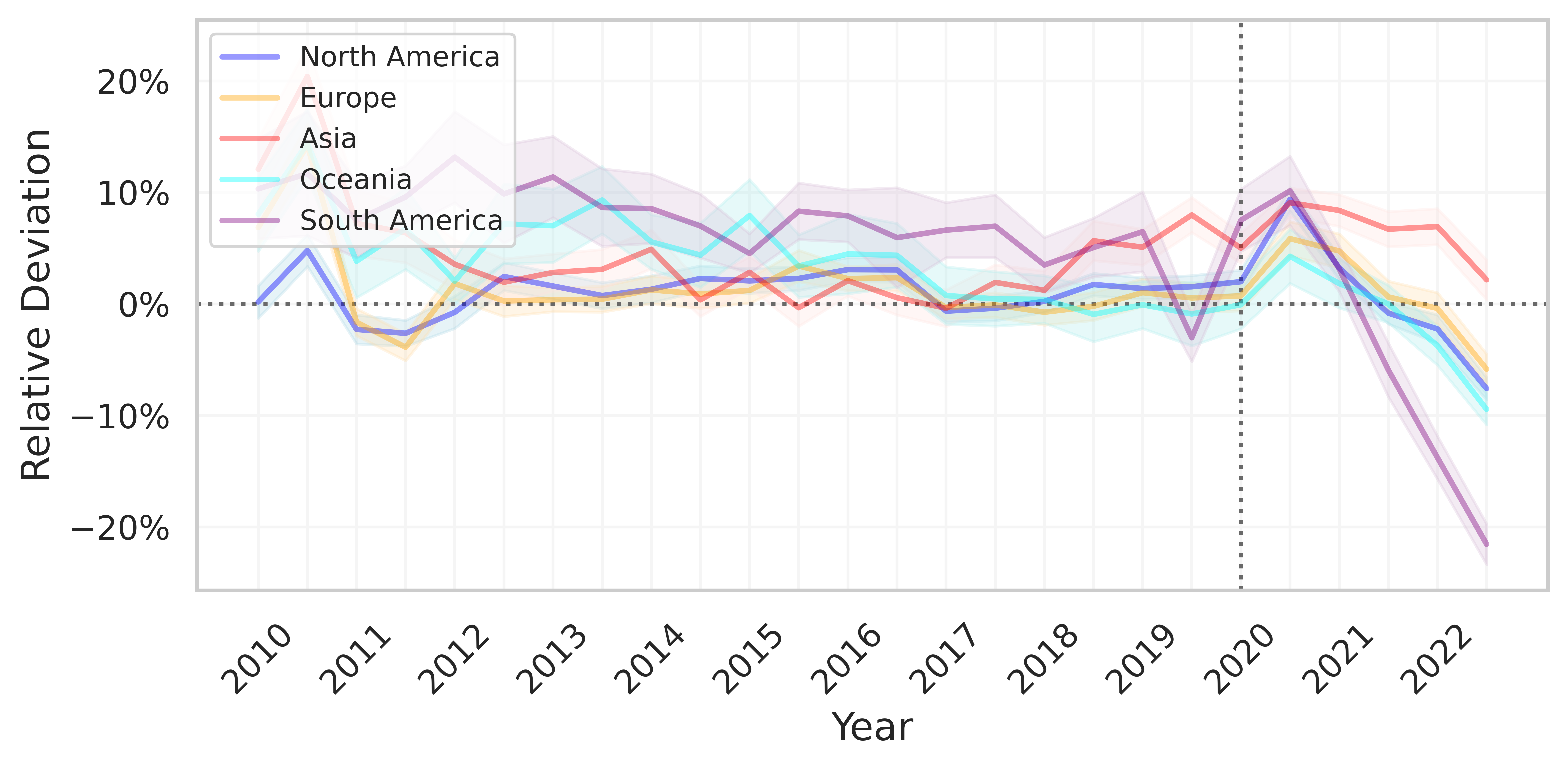}
    \end{minipage}
\end{figure}

\begin{figure}[bpth!]
    \centering
    \begin{minipage}{0.35\textwidth}
        \caption{\textit{Individual Productivity}, relative to 2010, in the top 1,000 universities grouped by \textit{Institutional Region}, calculated and plotted in six-month intervals between 2010-2022.}
        \label{fig:continent_productivity}
    \end{minipage}%
    \hfill
    \begin{minipage}{0.63\textwidth}
        \includegraphics[width=\textwidth]{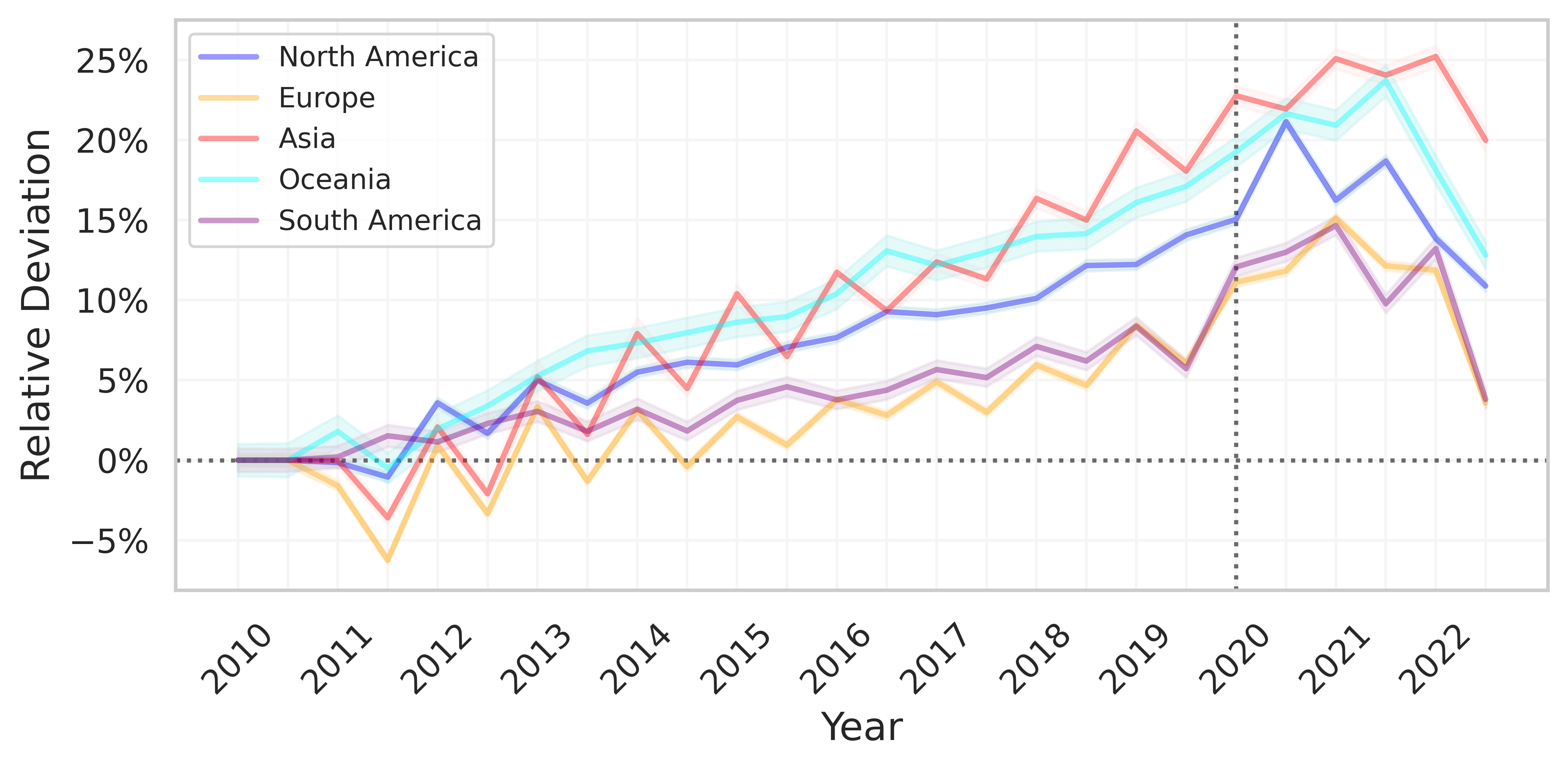}
    \end{minipage}
\end{figure}


\begin{table}[bpth!]
\centering
\resizebox{\textwidth}{!}{
\begin{tabular}{lcccccc}
\toprule
\textbf{Continent} & \textbf{2020 H1} & \textbf{2020 H2} & \textbf{2021 H1} & \textbf{2021 H2} & \textbf{2022 H1} & \textbf{2022 H2} \\
\midrule
\textbf{North America} &    $0.02\pm0.08$ &    $0.09\pm0.09$ &    $0.04\pm0.11$ &    $0.03\pm0.10$ &    $0.00\pm0.14$ &   $-0.03\pm0.11$ \\
\textbf{Europe       } &    $0.01\pm0.13$ &    $0.06\pm0.15$ &    $0.06\pm0.18$ &    $0.05\pm0.17$ &    $0.03\pm0.20$ &   $-0.01\pm0.18$ \\
\textbf{Africa       } &    $0.08\pm0.16$ &    $0.12\pm0.12$ &    $0.15\pm0.12$ &    $0.14\pm0.18$ &    $0.11\pm0.12$ &    $0.06\pm0.17$ \\
\textbf{Asia         } &    $0.05\pm0.11$ &    $0.09\pm0.10$ &    $0.10\pm0.13$ &    $0.11\pm0.15$ &    $0.13\pm0.20$ &    $0.11\pm0.21$ \\
\textbf{Oceania      } &   $-0.00\pm0.06$ &    $0.04\pm0.07$ &    $0.02\pm0.08$ &    $0.02\pm0.07$ &   $-0.03\pm0.08$ &   $-0.07\pm0.07$ \\
\textbf{South America} &    $0.08\pm0.08$ &    $0.10\pm0.09$ &    $0.07\pm0.08$ &   $-0.02\pm0.10$ &   $-0.09\pm0.08$ &   $-0.19\pm0.09$ \\
\bottomrule
\end{tabular}
}
\caption{This figure illustrates \textit{Author Participation} across the top 1,000 research institutions, categorized by \textit{Institutional Region}. It displays the average (and confidence interval) relative deviations in 2020 through 2022 based on a linear model trained using data up to the end of 2019.}
\label{tab:cont:auth}
\end{table}


\begin{table}[bpth!]
\centering
\resizebox{\textwidth}{!}{
\begin{tabular}{lcccccc}
\toprule
\textbf{Continent} & \textbf{2020 H1} & \textbf{2020 H2} & \textbf{2021 H1} & \textbf{2021 H2} & \textbf{2022 H1} & \textbf{2022 H2} \\
\midrule
\textbf{North America} &    $0.04\pm0.08$ &    $0.14\pm0.10$ &    $0.08\pm0.12$ &    $0.04\pm0.10$ &   $-0.01\pm0.13$ &   $-0.08\pm0.11$ \\
\textbf{Europe       } &    $0.05\pm0.14$ &    $0.11\pm0.16$ &    $0.13\pm0.20$ &    $0.07\pm0.19$ &    $0.06\pm0.21$ &   $-0.04\pm0.19$ \\
\textbf{Africa       } &    $0.12\pm0.11$ &    $0.19\pm0.15$ &    $0.25\pm0.23$ &    $0.21\pm0.20$ &    $0.21\pm0.20$ &    $0.13\pm0.27$ \\
\textbf{Asia         } &    $0.07\pm0.12$ &    $0.14\pm0.13$ &    $0.14\pm0.14$ &    $0.14\pm0.17$ &    $0.15\pm0.22$ &    $0.11\pm0.24$ \\
\textbf{Oceania      } &    $0.03\pm0.07$ &    $0.07\pm0.06$ &    $0.05\pm0.08$ &    $0.04\pm0.07$ &   $-0.02\pm0.08$ &   $-0.12\pm0.07$ \\
\textbf{South America} &    $0.12\pm0.09$ &    $0.16\pm0.09$ &    $0.14\pm0.08$ &    $0.01\pm0.09$ &   $-0.04\pm0.09$ &   $-0.20\pm0.09$ \\
\bottomrule
\end{tabular}
}
\caption{This figure illustrates \textit{Institutional Productivity} across the top 1,000 research institutions, categorized by \textit{Institutional Region}. It displays the average (and confidence interval) relative deviations in 2020 through 2022 based on a linear model trained using data up to the end of 2019.}
\label{tab:cont:work}
\end{table}

\begin{figure}[bpth!]
    \centering
    \begin{minipage}{0.35\textwidth}
        \caption{\textit{Average number of internal collaborations} (average node (author) degree)) in the top 1,000 universities grouped by \textit{Institutional Region}, 2010-2022. (a) Raw data: Average number of internal collaborations per institution, grouped by geographical region. (b) Relative deviation for average number of internal collaborations, trained on previous 10 years' data, calculated and plotted in six-month intervals.}
        \label{fig:continent_degree}
    \end{minipage}%
    \hfill
    \begin{minipage}{0.63\textwidth}
        \includegraphics[width=\textwidth]{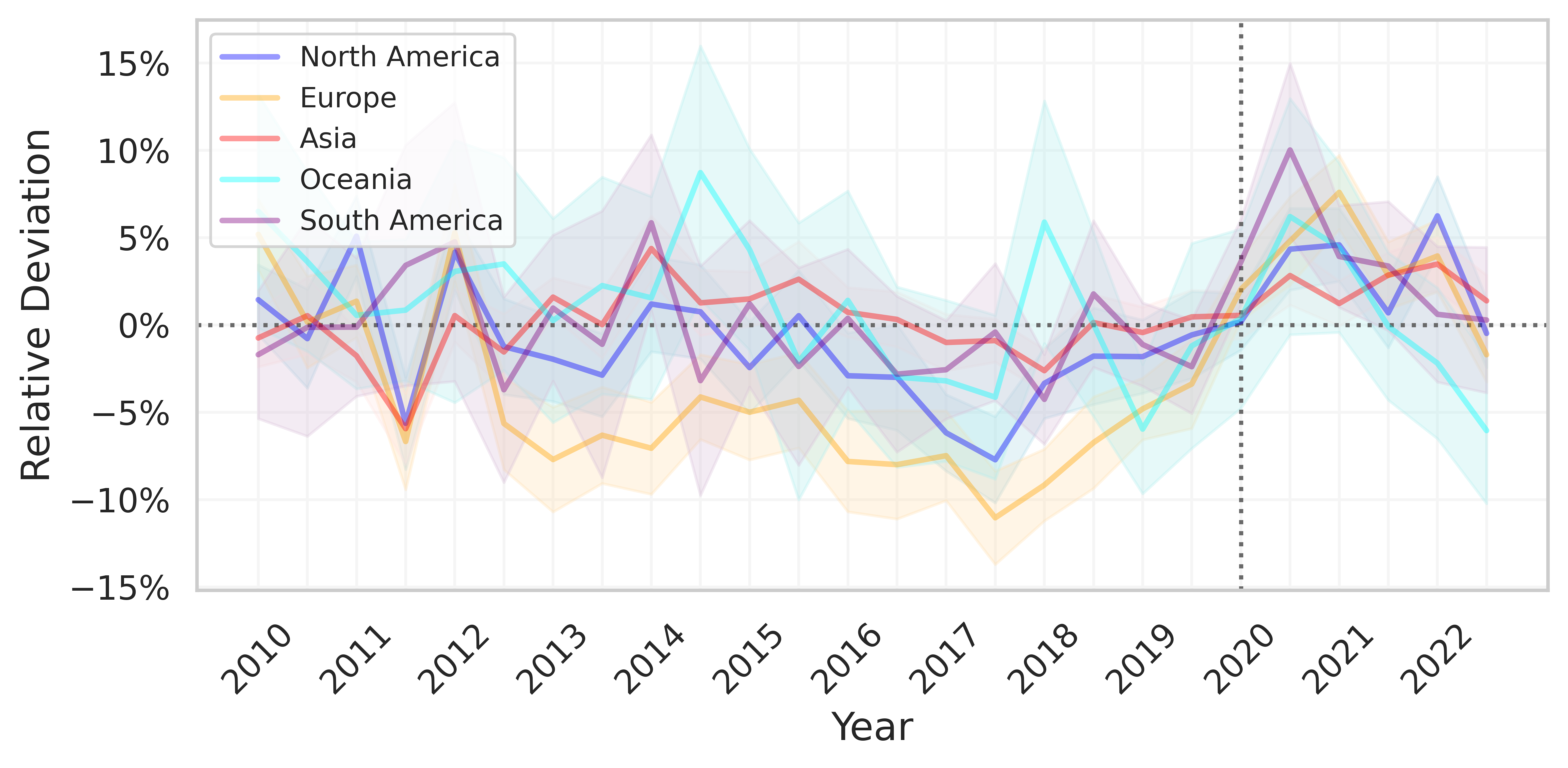}
    \end{minipage}
\end{figure}

\begin{figure}[bpth!]
    \centering
    \begin{minipage}{0.35\textwidth}
        \caption{\textit{Average institutional internal network clustering coefficient} at the top 1,000 universities grouped by \textit{Institutional Region}, 2010-2022. Relative deviation for average institutional internal network clustering coefficient, trained on previous 10 years' data, calculated and plotted in six-month intervals.}
        \label{fig:continent_clustercoef}
    \end{minipage}%
    \hfill
    \begin{minipage}{0.63\textwidth}
        \includegraphics[width=\textwidth]{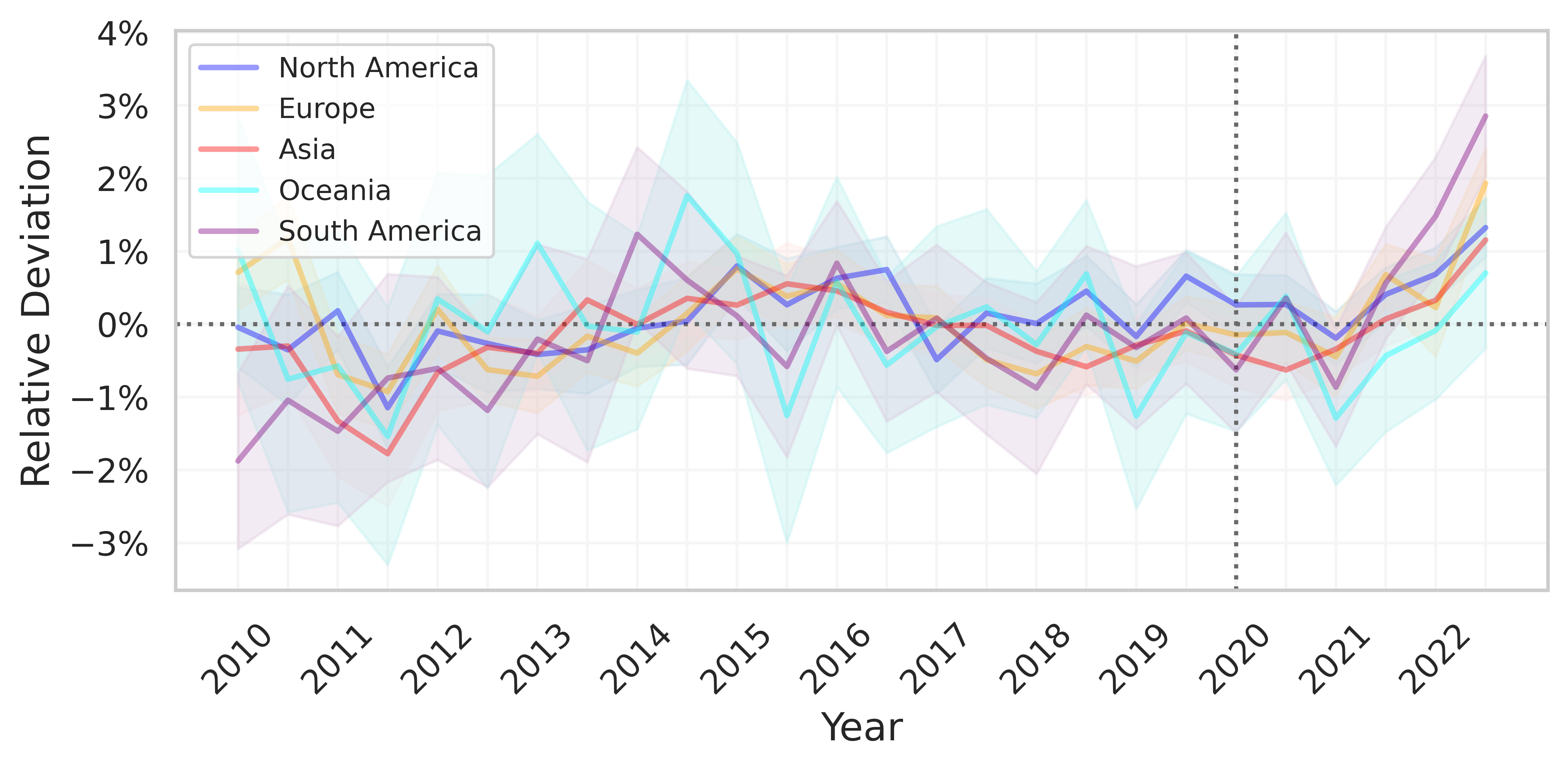}
    \end{minipage}
\end{figure}

\begin{figure}[bpth!]
    \centering
    \begin{minipage}{0.35\textwidth}
        \caption{\textit{Average internal team size} at the top 1,000 universities grouped by \textit{Institutional Region}, 2010-2022. Relative deviation for average team size, trained on previous 10 years' data, calculated and plotted in six-month intervals.}
        \label{fig:continent_teamsize}
    \end{minipage}%
    \hfill
    \begin{minipage}{0.63\textwidth}
        \includegraphics[width=\textwidth]{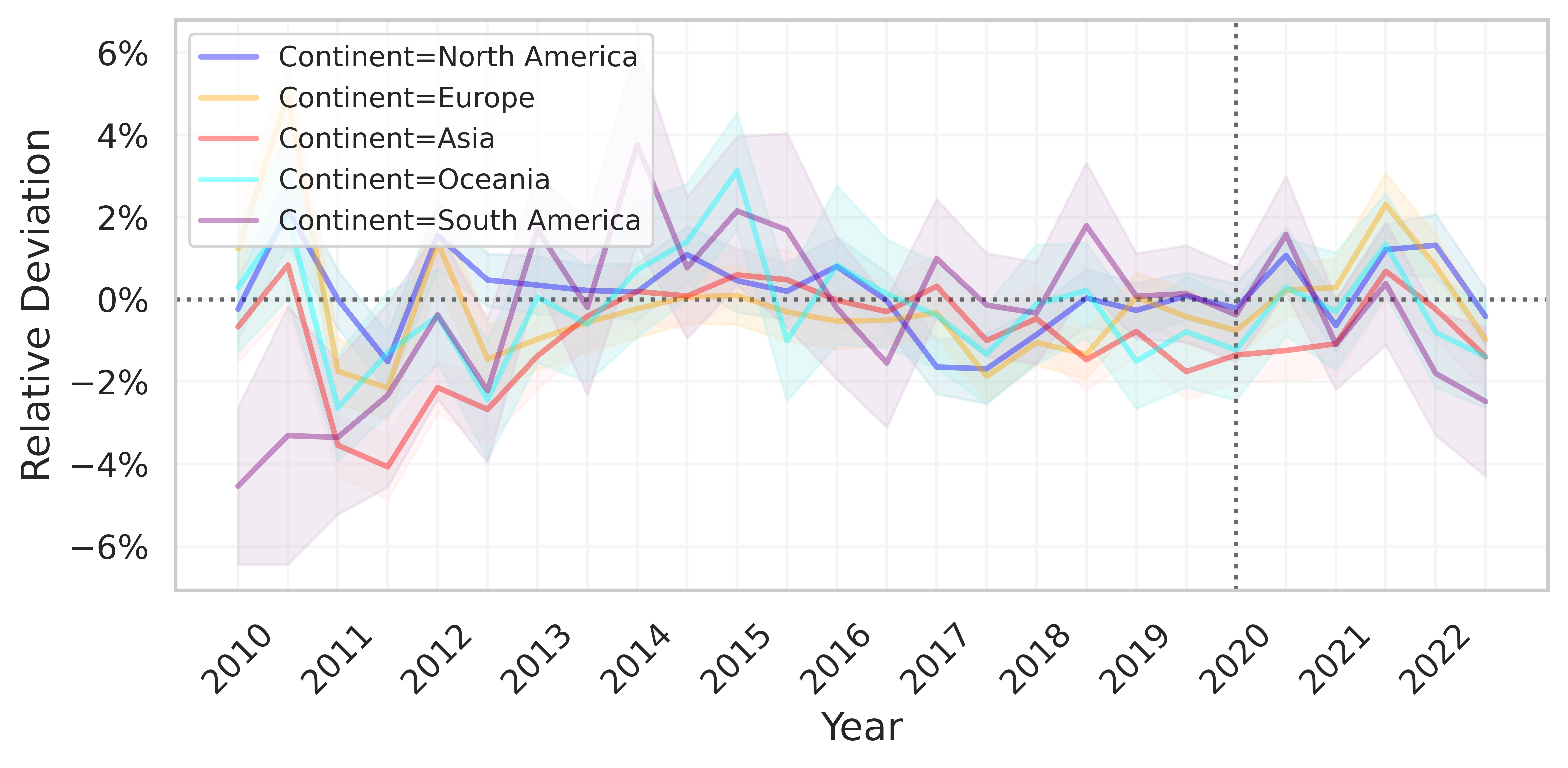}
    \end{minipage}
\end{figure}

\newpage
\section{Author Seniority}\label{secA3}

\begin{figure}[bpth!]
    \centering
    \begin{minipage}{0.35\textwidth}
        \caption{\textit{Institutional Productivity} in the top 1,000 universities grouped by \textit{Author Seniority}, 2010-2022. Relative deviation for average publications, trained on previous 10 years' data, calculated and plotted in six-month intervals.}
        \label{fig:seniority_publications}
    \end{minipage}%
    \hfill
    \begin{minipage}{0.63\textwidth}
        \includegraphics[width=\textwidth]{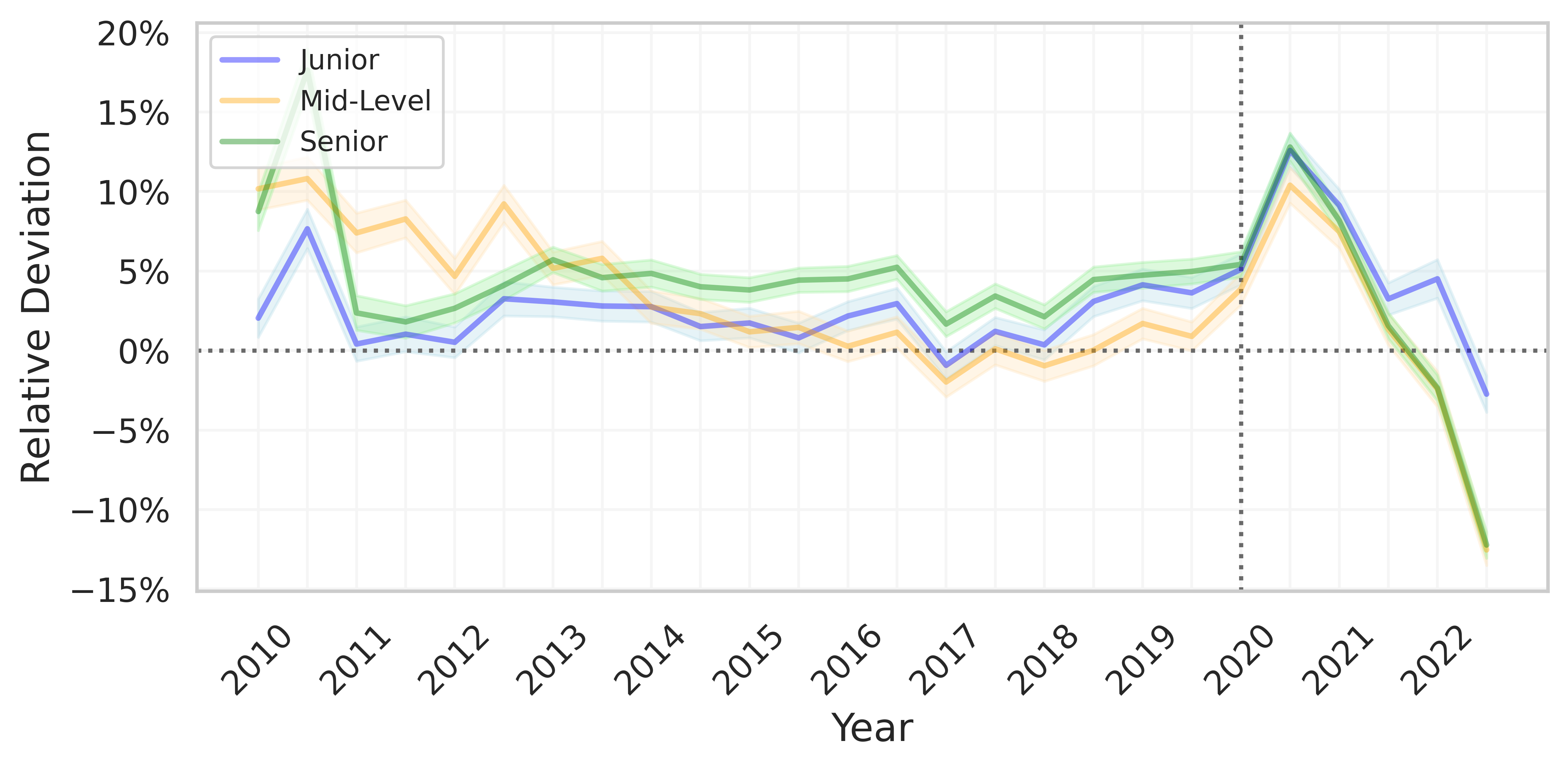}
    \end{minipage}
\end{figure}

\begin{figure}[bpth!]
    \centering
    \begin{minipage}{0.35\textwidth}
        \caption{\textit{Individual Productivity}, relative to 2010, in the top 1,000 universities grouped by \textit{Author seniority}, calculated and plotted in six-month intervals between 2010-2022.}
        \label{fig:seniority_productivity}
    \end{minipage}%
    \hfill
    \begin{minipage}{0.63\textwidth}
        \includegraphics[width=\textwidth]{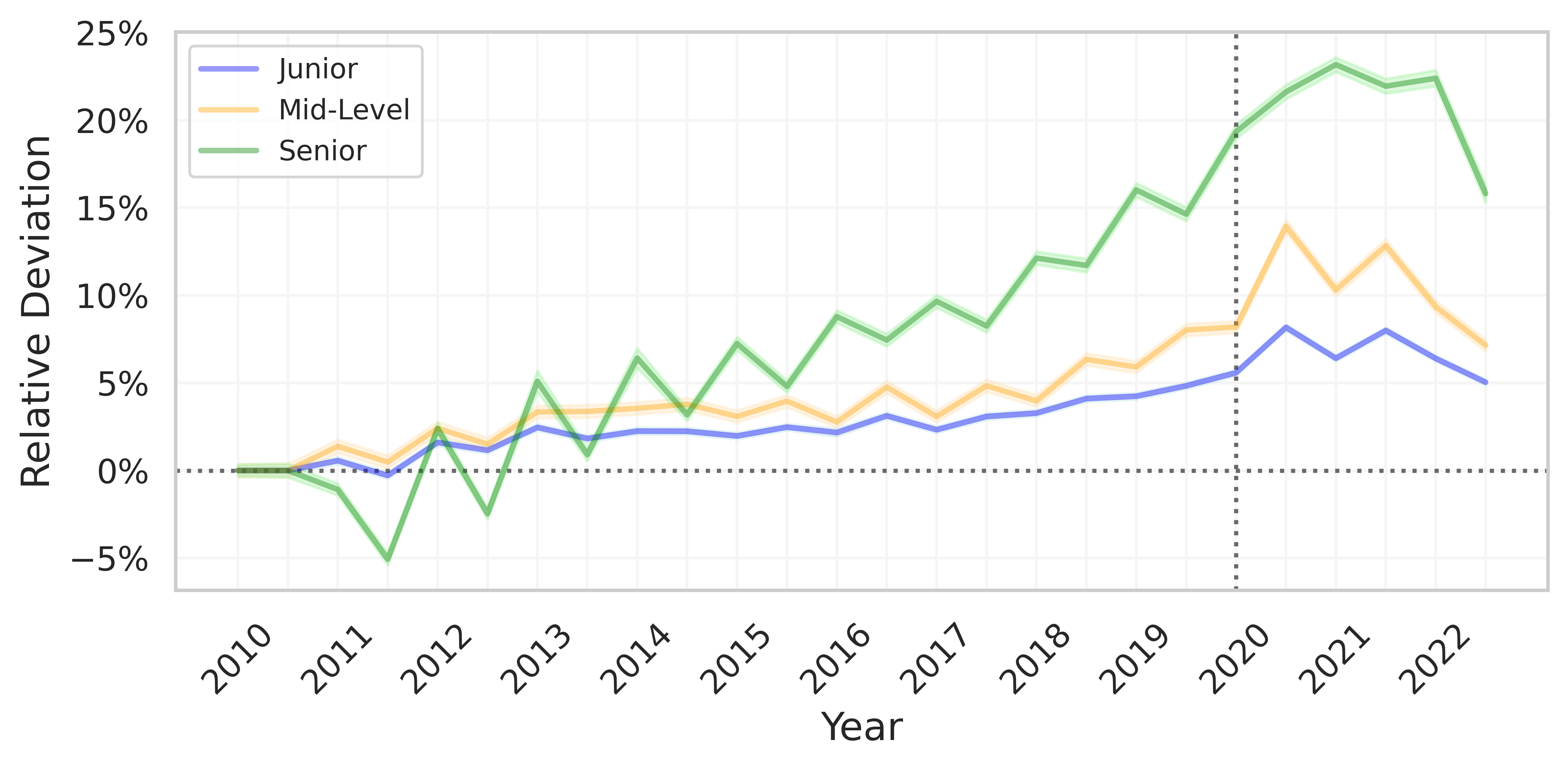}
    \end{minipage}
\end{figure}


\begin{table}[bpth!]
\centering
\resizebox{\textwidth}{!}{
\begin{tabular}{lcccccc}
\toprule
\textbf{Seniority} & \textbf{2020 H1} & \textbf{2020 H2} & \textbf{2021 H1} & \textbf{2021 H2} & \textbf{2022 H1} & \textbf{2022 H2} \\
\midrule
\textbf{Junior   } &    $0.04\pm0.14$ &    $0.11\pm0.16$ &    $0.10\pm0.19$ &    $0.08\pm0.19$ &    $0.13\pm0.28$ &    $0.10\pm0.28$ \\
\textbf{Mid-Level} &    $0.01\pm0.13$ &    $0.05\pm0.14$ &    $0.04\pm0.17$ &    $0.02\pm0.17$ &   $-0.02\pm0.20$ &   $-0.08\pm0.18$ \\
\textbf{Senior   } &    $0.02\pm0.11$ &    $0.08\pm0.12$ &    $0.06\pm0.14$ &    $0.06\pm0.15$ &    $0.02\pm0.17$ &   $-0.02\pm0.17$ \\
\bottomrule
\end{tabular}
}
\caption{This figure illustrates \textit{Author Participation} across the top 1,000 research institutions, categorized by \textit{Author Seniority}. It displays the average (and confidence interval) relative deviations in 2020 through 2022 based on a linear model trained using data up to the end of 2019.}
\label{tab:sen:auth}
\end{table}


\begin{table}[bpth!]
\centering
\resizebox{\textwidth}{!}{
\begin{tabular}{lcccccc}
\toprule
\textbf{Seniority} & \textbf{2020 H1} & \textbf{2020 H2} & \textbf{2021 H1} & \textbf{2021 H2} & \textbf{2022 H1} & \textbf{2022 H2} \\
\midrule
\textbf{Junior   } &    $0.05\pm0.15$ &    $0.12\pm0.17$ &    $0.11\pm0.19$ &    $0.09\pm0.20$ &    $0.10\pm0.26$ &    $0.05\pm0.27$ \\
\textbf{Mid-Level} &    $0.04\pm0.16$ &    $0.10\pm0.18$ &    $0.09\pm0.20$ &    $0.05\pm0.19$ &    $0.02\pm0.23$ &   $-0.08\pm0.21$ \\
\textbf{Senior   } &    $0.05\pm0.12$ &    $0.13\pm0.14$ &    $0.11\pm0.17$ &    $0.08\pm0.16$ &    $0.04\pm0.19$ &   $-0.05\pm0.19$ \\
\bottomrule
\end{tabular}
}
\caption{This figure illustrates \textit{Institutional Productivity} across the top 1,000 research institutions, categorized by \textit{Author Seniority}. It displays the average (and confidence interval) relative deviations in 2020 through 2022 based on a linear model trained using data up to the end of 2019.}
\label{tab:sen:work}
\end{table}

\begin{figure}[bpth!]
    \centering
    \begin{minipage}{0.35\textwidth}
        \caption{\textit{Average number of internal collaborations} (average node (author) degree)) in the top 1,000 universities grouped by \textit{Author seniority}, 2010-2022. (a) Raw data: Average number of internal collaborations per institution, grouped by geographical region. (b) Relative deviation for average number of internal collaborations, trained on previous 10 years' data, calculated and plotted in six-month intervals.}
        \label{fig:seniority_degree}
    \end{minipage}%
    \hfill
    \begin{minipage}{0.63\textwidth}
        \includegraphics[width=\textwidth]{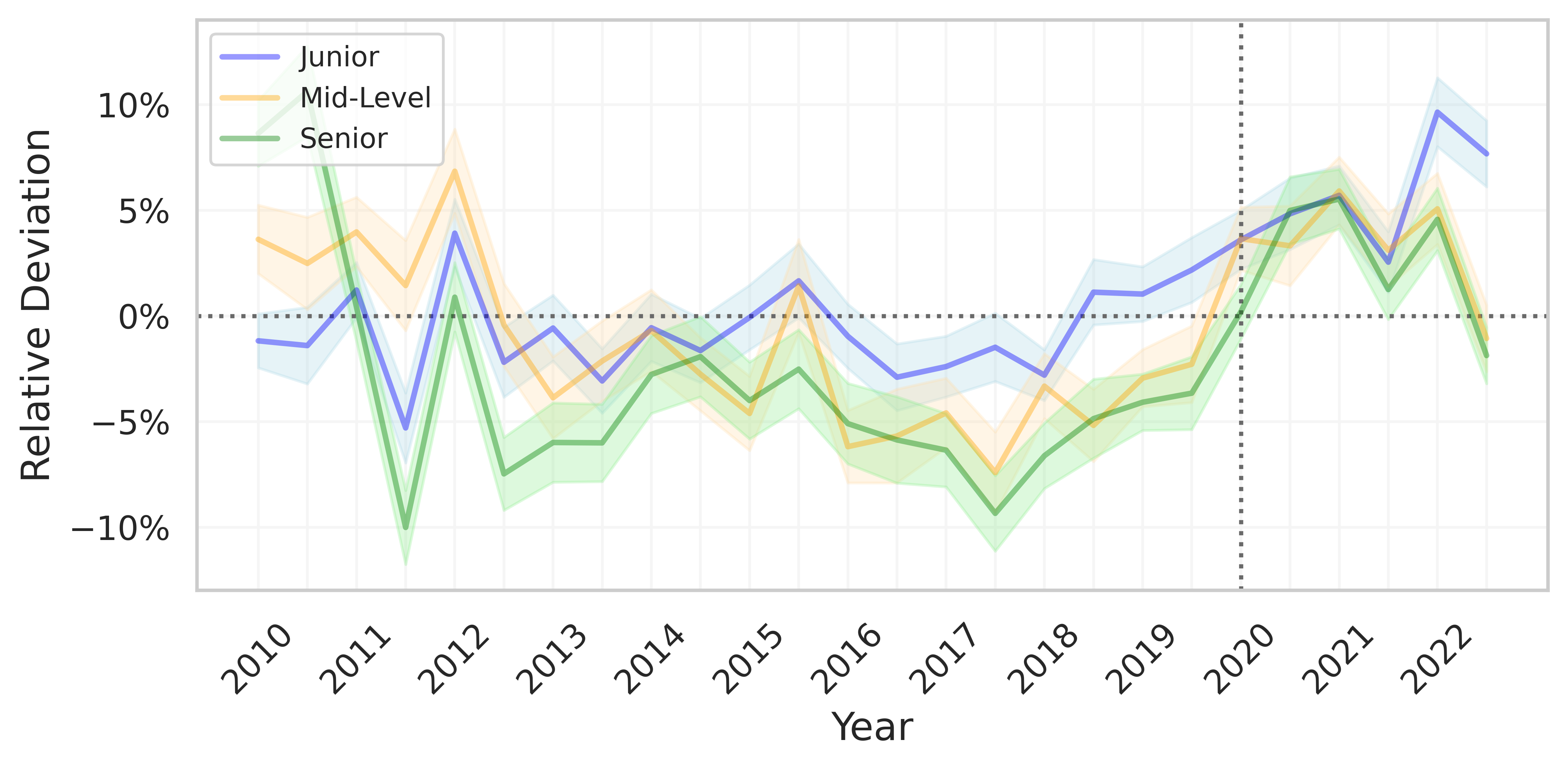}
    \end{minipage}
\end{figure}

\begin{figure}[bpth!]
    \centering
    \begin{minipage}{0.35\textwidth}
        \caption{\textit{Average internal team size} at the top 1,000 universities grouped by \textit{Author seniority}, 2010-2022. Relative deviation for average team size, trained on previous 10 years' data, calculated and plotted in six-month intervals.}
        \label{fig:seniority_teamsize}
    \end{minipage}%
    \hfill
    \begin{minipage}{0.63\textwidth}
        \includegraphics[width=\textwidth]{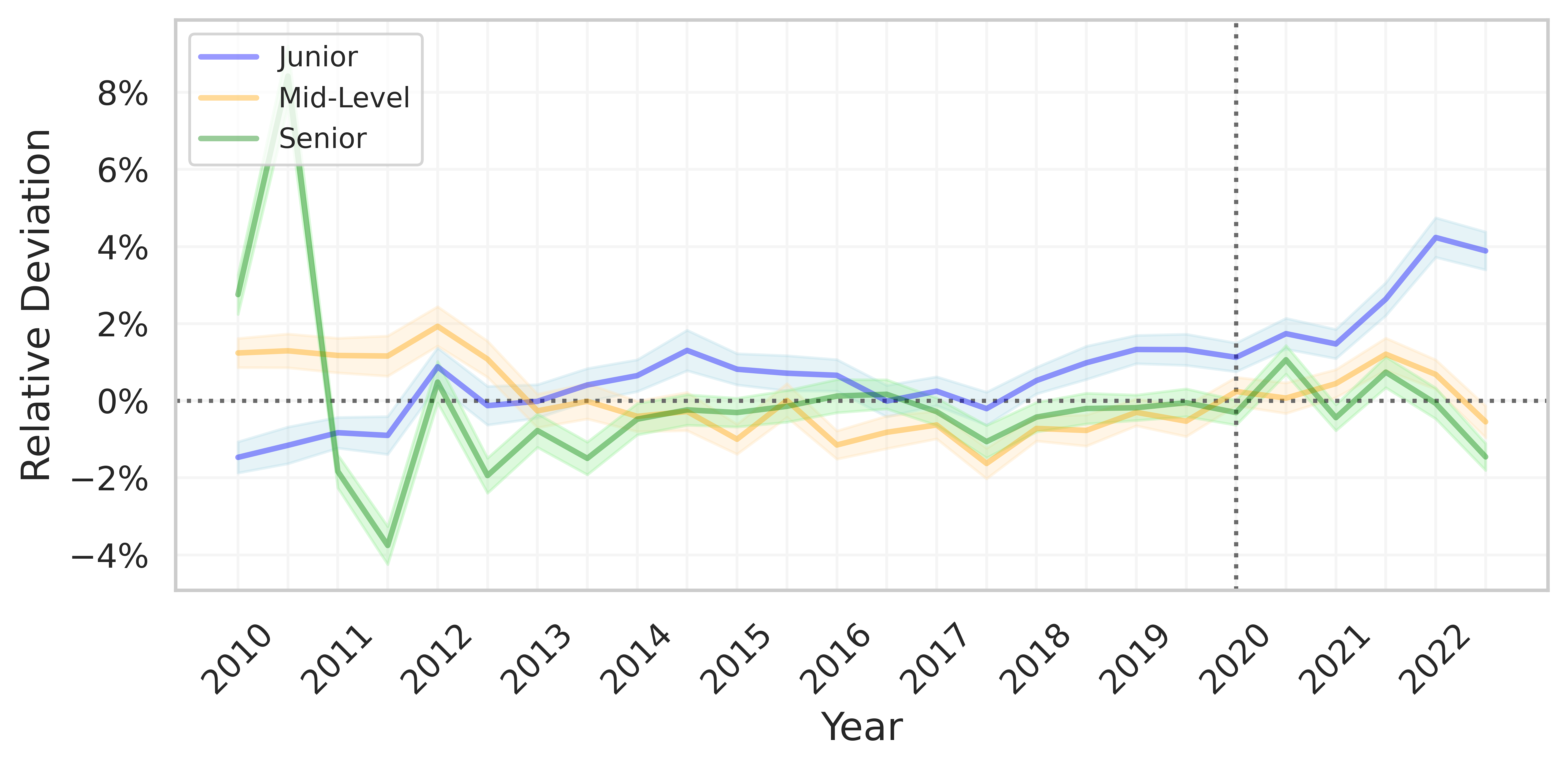}
    \end{minipage}
\end{figure}

\newpage
\section{Author Gender}\label{secA4}

\begin{figure}[bpth!]
    \centering
    \begin{minipage}{0.35\textwidth}
        \caption{\textit{Institutional Productivity} in the top 1,000 universities grouped by \textit{Author Gender}, 2010-2022. Relative deviation for average publications, trained on previous 10 years' data, calculated and plotted in six-month intervals.}
        \label{fig:gender_publications}
    \end{minipage}%
    \hfill
    \begin{minipage}{0.63\textwidth}
        \includegraphics[width=\textwidth]{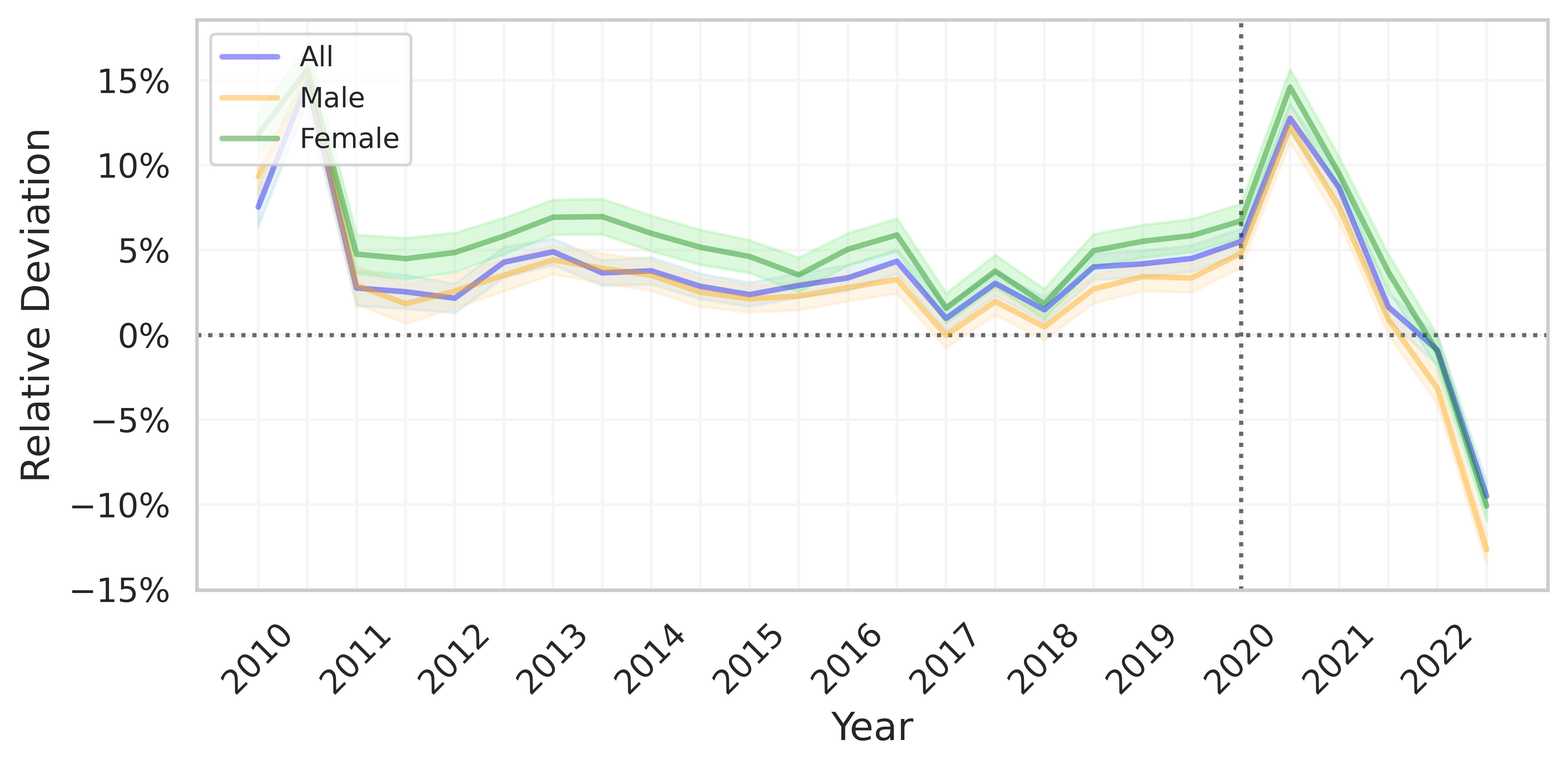}
    \end{minipage}
\end{figure}

\begin{figure}[bpth!]
    \centering
    \begin{minipage}{0.35\textwidth}
        \caption{\textit{Individual Productivity}, relative to 2010, in the top 1,000 universities grouped by \textit{Author gender}, calculated and plotted in six-month intervals between 2010-2022.}
        \label{fig:gender_productivity}
    \end{minipage}%
    \hfill
    \begin{minipage}{0.63\textwidth}
        \includegraphics[width=\textwidth]{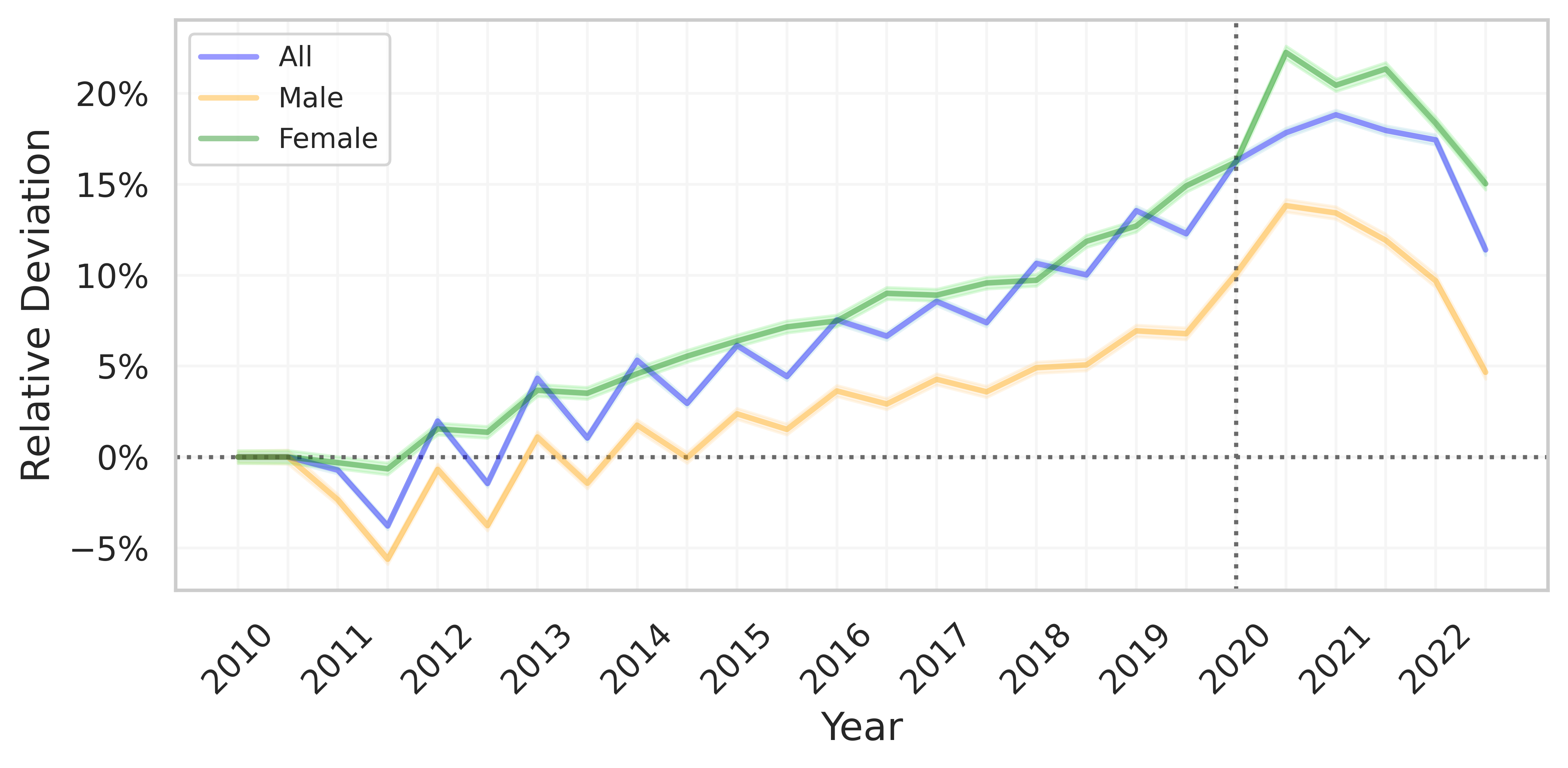}
    \end{minipage}
\end{figure}


\begin{table}[bpth!]
\centering
\resizebox{\textwidth}{!}{
\begin{tabular}{lcccccc}
\toprule
\textbf{Gender} & \textbf{2020 H1} & \textbf{2020 H2} & \textbf{2021 H1} & \textbf{2021 H2} & \textbf{2022 H1} & \textbf{2022 H2} \\
\midrule
\textbf{All   } &    $0.02\pm0.11$ &    $0.08\pm0.12$ &    $0.07\pm0.14$ &    $0.06\pm0.15$ &    $0.04\pm0.18$ &    $0.00\pm0.18$ \\
\textbf{Male  } &    $0.01\pm0.11$ &    $0.06\pm0.12$ &    $0.04\pm0.14$ &    $0.03\pm0.15$ &   $-0.01\pm0.18$ &   $-0.05\pm0.17$ \\
\textbf{Female} &    $0.04\pm0.14$ &    $0.10\pm0.15$ &    $0.08\pm0.16$ &    $0.08\pm0.17$ &    $0.05\pm0.20$ &    $0.01\pm0.19$ \\
\bottomrule
\end{tabular}
}
\caption{This figure illustrates \textit{Author Participation} across the top 1,000 research institutions, categorized by \textit{Author Gender}. It displays the average (and confidence interval) relative deviations in 2020 through 2022 based on a linear model trained using data up to the end of 2019.}
\label{tab:gen:auth}
\end{table}


\begin{table}[bpth!]
\centering
\resizebox{\textwidth}{!}{
\begin{tabular}{lcccccc}
\toprule
\textbf{Gender} & \textbf{2020 H1} & \textbf{2020 H2} & \textbf{2021 H1} & \textbf{2021 H2} & \textbf{2022 H1} & \textbf{2022 H2} \\
\midrule
\textbf{All   } &    $0.05\pm0.12$ &    $0.13\pm0.13$ &    $0.12\pm0.16$ &    $0.08\pm0.16$ &    $0.06\pm0.20$ &   $-0.02\pm0.20$ \\
\textbf{Male  } &    $0.05\pm0.13$ &    $0.12\pm0.14$ &    $0.10\pm0.17$ &    $0.07\pm0.17$ &    $0.02\pm0.20$ &   $-0.06\pm0.20$ \\
\textbf{Female} &    $0.07\pm0.16$ &    $0.14\pm0.16$ &    $0.13\pm0.19$ &    $0.10\pm0.19$ &    $0.07\pm0.23$ &   $-0.02\pm0.21$ \\
\bottomrule
\end{tabular}
}
\caption{This figure illustrates \textit{Institutional Productivity} across the top 1,000 research institutions, categorized by \textit{Author Gender}. It displays the average (and confidence interval) relative deviations in 2020 through 2022 based on a linear model trained using data up to the end of 2019.}
\label{tab:gen:work}
\end{table}

\newpage
\section{Fields of Study}\label{secA5}

\begin{figure}[bpth!]
    \centering
        \includegraphics[width=\textwidth]{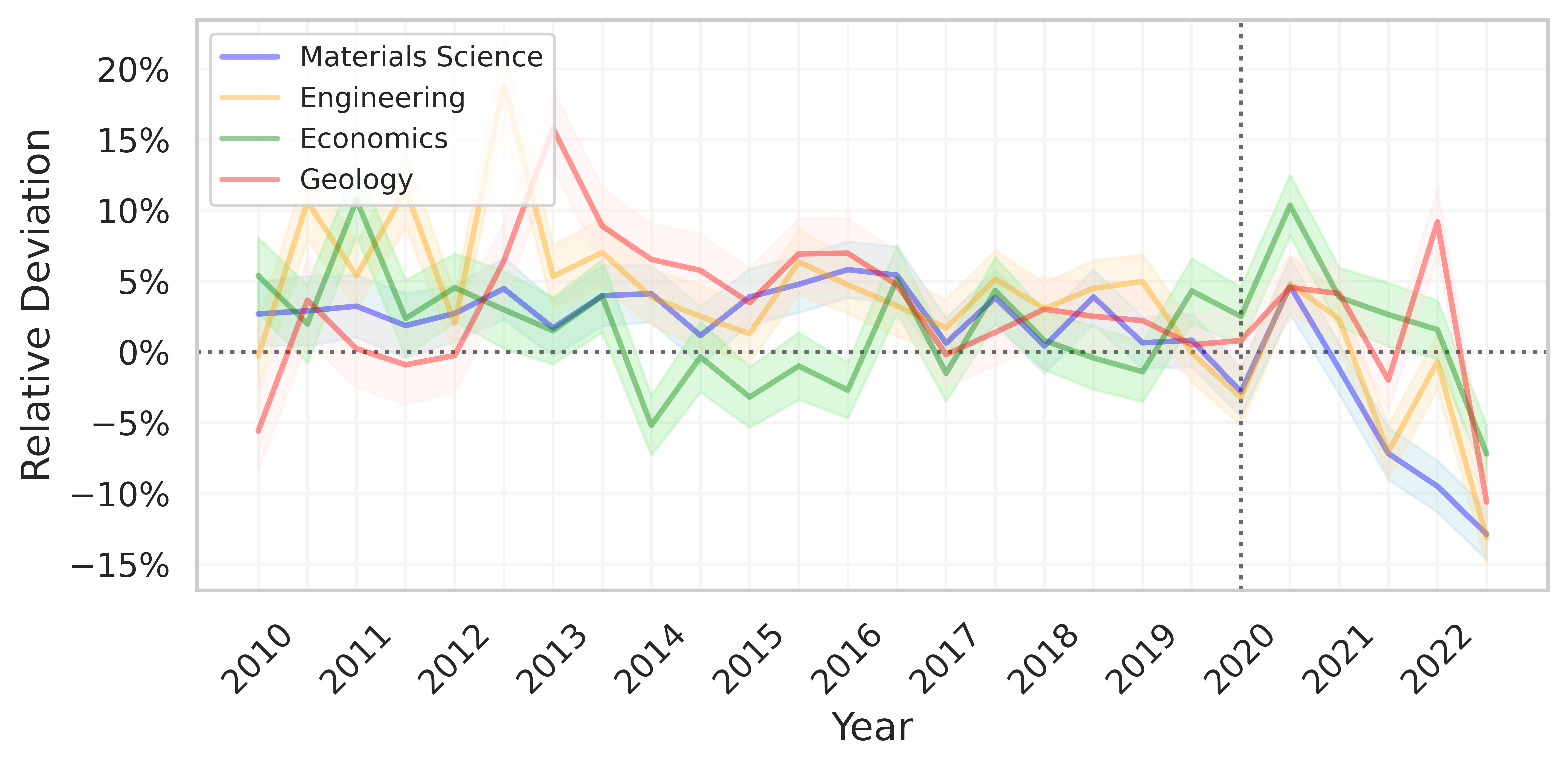}
        \caption{\textit{Institutional Productivity} in the top 1,000 universities grouped by \textit{Field of Study}, 2010-2022. Relative deviation for average publications, trained on previous 10 years' data, calculated and plotted in six-month intervals.}
        \label{fig:concept_publications2}
\end{figure}

\begin{figure}[bpth!]
    \centering
        \includegraphics[width=\textwidth]{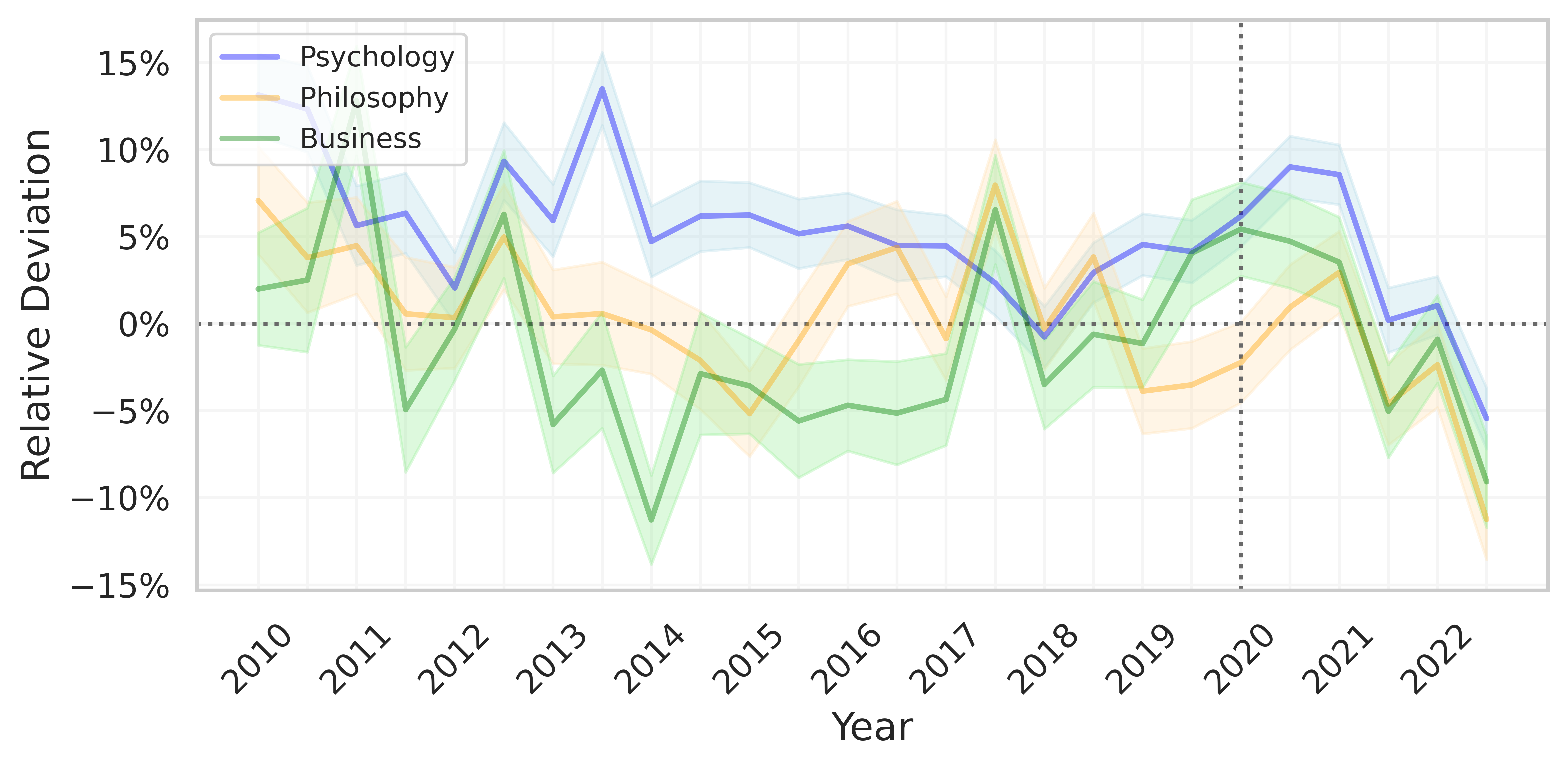}
        \caption{\textit{Institutional Productivity} in the top 1,000 universities grouped by \textit{Field of Study}, 2010-2022. Relative deviation for average publications, trained on previous 10 years' data, calculated and plotted in six-month intervals.}
        \label{fig:concept_publications3}
\end{figure}


\begin{table}[bpth!]
\centering
\resizebox{\textwidth}{!}{
\begin{tabular}{lcccccc}
\toprule
\textbf{Concept} & \textbf{2020 H1} & \textbf{2020 H2} & \textbf{2021 H1} & \textbf{2021 H2} & \textbf{2022 H1} & \textbf{2022 H2} \\
\midrule
\textbf{Biology              } &    $0.08\pm0.17$ &    $0.12\pm0.18$ &    $0.16\pm0.21$ &    $0.09\pm0.20$ &    $0.08\pm0.25$ &   $-0.00\pm0.24$ \\
\textbf{Medicine             } &    $0.14\pm0.20$ &    $0.27\pm0.22$ &    $0.25\pm0.26$ &    $0.19\pm0.24$ &    $0.14\pm0.26$ &    $0.04\pm0.26$ \\
\textbf{Chemistry            } &    $0.08\pm0.23$ &    $0.16\pm0.26$ &    $0.11\pm0.29$ &    $0.15\pm0.31$ &    $0.07\pm0.33$ &    $0.12\pm0.34$ \\
\textbf{Computer Science     } &    $0.08\pm0.23$ &    $0.09\pm0.23$ &    $0.16\pm0.27$ &    $0.10\pm0.27$ &    $0.17\pm0.32$ &    $0.03\pm0.31$ \\
\textbf{Physics              } &   $-0.03\pm0.22$ &    $0.02\pm0.23$ &    $0.01\pm0.24$ &   $-0.00\pm0.26$ &   $-0.04\pm0.27$ &   $-0.11\pm0.27$ \\
\textbf{Materials Science    } &   $-0.03\pm0.25$ &    $0.05\pm0.29$ &   $-0.03\pm0.28$ &   $-0.05\pm0.30$ &   $-0.11\pm0.31$ &   $-0.13\pm0.33$ \\
\textbf{Engineering          } &   $-0.05\pm0.29$ &    $0.04\pm0.28$ &    $0.01\pm0.31$ &   $-0.06\pm0.29$ &   $-0.02\pm0.33$ &   $-0.15\pm0.32$ \\
\textbf{Psychology           } &    $0.05\pm0.25$ &    $0.07\pm0.26$ &    $0.10\pm0.28$ &    $0.02\pm0.29$ &    $0.05\pm0.31$ &   $-0.04\pm0.31$ \\
\textbf{Economics            } &    $0.00\pm0.28$ &    $0.08\pm0.31$ &    $0.02\pm0.31$ &    $0.04\pm0.33$ &    $0.00\pm0.33$ &   $-0.06\pm0.34$ \\
\textbf{Philosophy           } &   $-0.03\pm0.32$ &    $0.01\pm0.33$ &    $0.02\pm0.34$ &   $-0.05\pm0.34$ &   $-0.03\pm0.37$ &   $-0.13\pm0.34$ \\
\textbf{Political Science    } &    $0.04\pm0.29$ &    $0.12\pm0.31$ &    $0.10\pm0.32$ &    $0.08\pm0.33$ &    $0.10\pm0.32$ &   $-0.01\pm0.33$ \\
\textbf{Sociology            } &    $0.01\pm0.33$ &    $0.08\pm0.34$ &    $0.08\pm0.35$ &    $0.04\pm0.34$ &    $0.05\pm0.36$ &   $-0.05\pm0.35$ \\
\textbf{Geology              } &   $-0.01\pm0.30$ &    $0.05\pm0.30$ &    $0.03\pm0.32$ &   $-0.00\pm0.32$ &    $0.08\pm0.36$ &   $-0.11\pm0.31$ \\
\textbf{Environmental Science} &   $-0.11\pm0.44$ &   $-0.08\pm0.44$ &   $-0.07\pm0.42$ &   $-0.15\pm0.42$ &   $-0.13\pm0.42$ &   $-0.26\pm0.39$ \\
\textbf{Business             } &    $0.02\pm0.36$ &    $0.04\pm0.35$ &    $0.02\pm0.36$ &   $-0.04\pm0.37$ &   $-0.01\pm0.40$ &   $-0.11\pm0.39$ \\
\textbf{History              } &   $-0.01\pm0.38$ &   $-0.05\pm0.39$ &   $-0.09\pm0.37$ &   $-0.14\pm0.37$ &   $-0.11\pm0.40$ &   $-0.23\pm0.35$ \\
\textbf{Art                  } &   $-0.02\pm0.38$ &   $-0.03\pm0.35$ &   $-0.04\pm0.35$ &   $-0.07\pm0.35$ &   $-0.12\pm0.38$ &   $-0.18\pm0.35$ \\
\textbf{Mathematics          } &   $-0.01\pm0.25$ &    $0.01\pm0.27$ &    $0.03\pm0.28$ &   $-0.02\pm0.28$ &   $-0.03\pm0.30$ &   $-0.11\pm0.28$ \\
\textbf{Geography            } &    $0.03\pm0.34$ &    $0.04\pm0.35$ &    $0.05\pm0.34$ &   $-0.06\pm0.33$ &    $0.01\pm0.38$ &   $-0.15\pm0.36$ \\
\bottomrule
\end{tabular}
}
\caption{This figure illustrates \textit{Institutional Productivity} across the top 1,000 research institutions, categorized by \textit{Field of Study}. It displays the average (and confidence interval) relative deviations in 2020 through 2022 based on a linear model trained using data up to the end of 2019.}
\label{tab:conc:work}
\end{table}

\end{appendices}

\clearpage 

\bibliography{sn-bibliography}

\end{document}